\documentclass{article}

\usepackage{arxiv}

\usepackage[utf8]{inputenc} 
\usepackage[T1]{fontenc}    
\usepackage{hyperref}       
\usepackage{url}            
\usepackage{booktabs}       
\usepackage{amsfonts}       
\usepackage{nicefrac}       
\usepackage{microtype}      
\usepackage{lipsum}
\usepackage{siunitx}
\usepackage{amsmath}
\usepackage{caption}
\usepackage{subcaption}
\usepackage{xcolor}
\usepackage{graphicx}
\usepackage[superscript]{cite}
\usepackage{sidecap}
\usepackage{comment}
\usepackage{multirow}

\title{Solubility of organic salts in solvent-antisolvent mixtures: A combined experimental and molecular dynamics simulations approach}

\author{
  Zoran ~Bjelobrk\\
  Institute of Energy and Process Engineering\\
  ETH Z\"urich, CH-8092, Switzerland\\
  \And
  Ashwin ~Kumar ~Rajagopalan\\
  Department of Chemical Engineering\\
  University of Manchester\\
  Manchester, M13 9PL, United Kingdom\\
  \And
  Dan ~Mendels\\
  Pritzker School of Molecular Engineering\\
  University of Chicago\\
  Chicago, Illinois 60637, United States\\
  \And
 Tarak ~Karmakar\\
 Department of Chemistry\\
  Indian Institute of Technology, Delhi\\
  Hauz Khas, New Delhi 110016, India\\
  \And
  Michele ~Parrinello\thanks{michele.parrinello@iit.it}\\
  Istituto Italiano di Tecnologia (IIT)\\
  Via Morego, 30, 16163 Genova GE, Italy\\
  \And
  Marco ~Mazzotti\thanks{marco.mazzotti@ipe.mavt.ethz.ch}\\
  Institute of Energy and Process Engineering\\
  ETH Z\"urich, CH-8092, Switzerland\\
}

\begin{document}
\maketitle

\begin{abstract}

We combine molecular dynamics simulations with experiments to estimate solubilities of organic salts in complex growth environments.
We predict the solubility by simulations of the growth and dissolution of ions at the crystal surface kink sites at different solution concentrations. Thereby, the solubility is identified as the solution's salt concentration, where the energy of the ion pair dissolved in solution equals the energy of the ion pair crystallized at the kink sites.
The simulation methodology is demonstrated for the case of anhydrous sodium acetate crystallized from various solvent-antisolvent mixtures.
To validate the predicted solubilities, we have measured the solubilities of sodium acetate in-house, using an experimental setup and measurement protocol that guarantees moisture-free conditions, which is key for a hygroscopic compound like sodium acetate.
We observe excellent agreement between the experimental and the computationally evaluated solubilities for sodium acetate in different solvent-antisolvent mixtures.
Given the agreement and the rich data the simulations produce, we can use them to complement experimental tasks which in turn will reduce time and capital in the design of complicated industrial crystallization processes of organic salts.

\end{abstract}

\section{Introduction}

Solubilities of organic salts play an important role in the pharmaceutical industry. Salt formation is a common way of tailoring the solubility of an active pharmaceutical ingredient (API) thus modifying its dissolution rate\cite{Berge1977, Paulekuhn2007}. 
Typically, in the design and development of the drug's crystallization process, the API's solubility is the first property that is quantified via experimental measurements. However, with the improving capabilities of computational methods, and in particular molecular dynamics (MD) simulations, future measurements and high throughput studies both in academia and in the industry will benefit tremendously from an experimentally validated computational approach.
Here, we demonstrate that this is possible by combining  a dedicated experimental setup together with a recently introduced MD method\cite{Bjelobrk2021}. We find that such a framework yields an accurate estimation of key properties like solubility, thus paving the way for a combined methodology, which has the potential of significantly reducing time and the cost of such an endeavour.

Most often, to measure the solubility of a given compound in crystallization processes, one employs gravimetric, spectroscopic, and chromatographic experiments, or a combination thereof.\cite{Togkalidou2002,Hu2005,Cornel2008,Borissova2008,Reus2015,Simone2015,Yang2016} Despite their widespread use, these methods  suffer from several technical challenges, which are subject of active research\cite{Botschi2018}. In hygroscopic or thermally unstable compounds, all these methods fail if adopted in the absence of careful consideration of the experimental environment. For example, to maintain a moisture-free environment one might have to perform experiments in an inert atmosphere, e.g., in a glove-box or in a Schlenk line. For compounds with low solubility, low spectral peak sensitivity, or low absolute changes in concentration, spectroscopic methods do not yield accurate concentration estimates. Even though some of these issues can be resolved through chromatography, they usually require a tedious and time-consuming sample preparation step. It must be kept in mind that even despite being the norm, the experimental approach to measure solubility is not efficient in terms of resources and time required as it will be evident from the experimental study presented below.

MD simulations have become an effective tool to gain insights into crystallization phenomena and to resolve them at the atomistic level\cite{Stack2012,Salvalaglio2012,DeLaPierre2017,Joswiak2018a,Karmakar2019}.
In the case of organic salts, MD simulations can be used to predict solubility and to understand the mechanism of ion attachment and detachment during crystal growth and dissolution, respectively.
As such, MD simulations help reducing empiricism, and guiding and accelerating the  experimental campaign  for crystallization process development.
Recently we have introduced an MD simulation setup that allows the prediction of solubility of organic molecules in various solvents.\cite{Bjelobrk2021}
This approach concentrates on the growth and dissolution process at kink sites, which constitute the end points of unfinished molecular (or ionic) rows on the edges of a crystal surface\cite{Kossel1927, Stranski1928, Burton1951}, as kink sites are the most relevant growth and dissolution sites for solutions at concentrations close to the solubility limit\cite{Snyder2007}.

In this work, we intend to show by direct comparison with experiments that this MD simulation approach, once properly tailored, can be used to  predict reliably the solubility of organic salts.
For the first time, we present a methodology to estimate the solubility of organic salts, by extending our previous work on kink growth and dissolution\cite{Bjelobrk2021}.
Note that the extension of this approach from organic crystals (one species involved in the growth and dissolution events) to organic salts where at least two ions must be considered is  challenging per se. In addition, since we consider different  solvent-antisolvent mixtures, one has to deal with a ternary system. In this work, we have selected anhydrous sodium acetate (NaOAc) and its polymorph I\cite{Hsu1983} as the organic salt on which to apply the new approach. We have used methanol (MeOH) as the solvent and either propan-1-ol (PrOH) or acetonitrile (MeCN) as the  antisolvent, in either  cases  mixtures with different solvent-antisolvent ratios were considered.

NaOAc's properties are well documented in the literature \cite{Lide1992,Soleymani2013,Kashefolgheta2017} and sodium is frequently used as a counterion in salt formulations of acidic APIs\cite{Paulekuhn2007}.
NaOAc is not only a suitable model compound to study  organic salts, but
also an important substance with a wide range of uses.
Most notably, its trihydrate form is an important phase-change material\cite{Kumar2017}, which can absorb and release heat through solid state changes. In the design of phase-change materials, it is important to understand the properties of all of its solid state forms, including the anhydrous ones \cite{Dittrich2018}.
NaOAc's solubility is weakly dependent on temperature, which is common for salts, and therefore the use of antisolvents is necessary to induce and control  crystallization. 
Despite having a simple form for an organic salt, NaOAc is a challenging system, both in experiments and in simulations.

Anhydrous NaOAc rapidly changes into its hydrated form upon adsorption of moisture as it inevitably does when it comes into contact with ambient atmosphere\cite{Cameron1976}. Therefore, carrying out experimental measurements with NaOAc suffer from challenges that are related to the adsorption of moisture. Thus, we undertook a laborious but successful experimental campaign, in which we took care of avoiding exposure to moisture. First, we handled the anhydrous NaOAc in a glove-box. Second, we performed all the solubility measurements in a Schlenk line. Finally, we used chromatography instead of a gravimetric method. We overcame the challenge and obtained the solubility curves for NaOAc in different solvent-antisolvent mixtures, which could be used as a reference to gauge the predictive capabilities of the MD simulations.

MD simulations have the advantage to provide more control over the crystal-solution system, but simulations of NaOAc crystal growth in solvent-antisolvent mixtures pose considerable hurdles as well, which manifest in size and time-scale limitations.
In regular MD simulations, as the crystal grows, the solution is quickly depleted due to system size limitations. Such depletion can be prevented with the constant chemical potential molecular dynamics (C$\mu$MD) algorithm, developed by Perego \emph{et al.}\cite{Perego2015}.
The NaOAc simulations require the control of the chemical potential of a complex solution comprised of two dissociated ions, a solvent, and an antisolvent. 
With the right choice of system parameters, the C$\mu$MD algorithm allowed us to keep the solution concentration constant in the proximity of the crystal surface.
Furthermore, salt compounds have large activation energy barriers both for growth and dissolution\cite{Stack2012,DeLaPierre2017,Joswiak2018a}.
Thus,  it is impossible to run an ordinary MD simulation  for a time long enough to observe a  number of growth and dissolution events sufficient to calculate solubility with any accuracy.  To overcome this limitation, we use well-tempered Metadynamics\cite{Barducci2008} (WTMetaD) together with a set of collective variables\cite{Bjelobrk2021} (CVs), which capture the slow degrees of freedom for the kink growth and dissolution for sodium (Na$^+$) and acetate (AcO$^-$) ions.

\section{Methodology}

\subsection{Experiments}
\label{sec:experiments}
We measure the solubility of NaOAc, $\chi_\text{exp}^*$, in different solvent-antisolvent mixtures using the equilibrium concentration method. In this method, we add excess salt to a given solvent-antisolvent mixture and let the solution equilibrate for a given period of time; we assume the concentration measured at the end of the equilibration time to be the solubility of the salt at that specific condition.

The experimental protocol consists of two steps, namely, a sample preparation step and a concentration estimation step. In the sample preparation step, we add excess anhydrous sodium acetate (NaOAc, anhydrous $\geq \SI{99}{\percent}$, Sigma Aldrich, Buchs, Switzerland) to a given mixture of solvent-antisolvent. We use methanol (MeOH, gradient grade, Merck KGaA, Darmstadt, Germany) as the solvent for all the experiments and we use propan-1-ol (PrOH, $\geq \SI{99.5}{\percent}$ Fisher Scientific, Reinach, Switzerland) or acetonitrile (MeCN, $\geq \SI{99.5}{\percent}$ Sigma Aldrich, Buchs, Switzerland) as the antisolvent. We prepare pure solvent (\num{100}\si{\percent} MeOH) as well as solvent-antisolvent mixtures (on a weight basis) by mixing predetermined quantities of MeOH-PrOH (\num{80}-\num{20}\si{\percent}, \num{60}-\num{40}\si{\percent}, \num{40}-\num{60}\si{\percent}) or MeOH-MeCN (\num{75}-\num{25}\si{\percent}, \num{50}-\num{50}\si{\percent}). To avoid atmospheric moisture exposure, we handle the NaOAc crystals in a glovebox under an argon environment. We add excess NaOAc crystals to a three neck \SI{500}{\milli\liter} round-bottom flask with a magnetic stirrer, we seal it and move it from the glovebox to a Schlenk line. We flush the Schlenk line and saturate it with argon, before opening the flask to add the solvent-antisolvent mixture prepared in advance. Note that we perform this addition using a syringe, with the flask attached to the Schlenk line and under a constant flow of argon. Upon the complete addition of the solvent-antisolvent mixture, we seal the flask and move it to a thermal bath, which is kept at a constant temperature of \SI{25}{\degreeCelsius}. We stir this suspension at constant temperature for \SI{20}{\hour}. Based on preliminary experiments to estimate the solubility, where the suspension was saturated for \SI{20}{\hour}, \SI{48}{\hour}, and \SI{72}{\hour}, we concluded that \SI{20}{\hour} was sufficient to guarantee equilibrium between the crystals of the solute and its solution in the solvent-antisolvent mixture.

For the concentration measurement, we use high performance liquid chromatography (HPLC), i.e., by measuring the amount of acetate present in a given sample. To this aim, we first obtain a solid-free saturated solution from the previous step. To obtain this solution, we reattach the flask with the suspension to the Schlenk line under an argon environment. We use a second three neck \SI{500}{\milli\liter} round-bottom flask to collect the solids-free saturated solution. These two flasks are connected using a Teflon tube with a filter element to remove the solids. We pull vacuum through the second flask attached to the Schlenk line and we exploit the pressure gradient between the two flasks to facilitate the transfer of suspension from the first flask through the filter to the second flask. Finally, we switch from a vacuum environment to an argon environment. Subsequently, we take samples of the filtered saturated solution and dilute it using deionized and filtered (filter size of \SI{0.22}{\micro\meter}) water obtained from a Milli-QAdvantage A10 system (Millipore, Zug, Switzerland). We perform this step as the HPLC is typically employed to detect concentrations under dilute conditions. To convert the HPLC chromatogram to a concentration estimate, we performed a thorough calibration using acetic acid as the reference (i.e., a calibration curve that relates area under the curve for the acetic acid peak in the chromatogram to the acetic acid concentration). In our actual experiments, upon dilution of the saturated solution in water, the sodium acetate dissociates into sodium and acetate ions. We use sulphuric acid (pH $= 2$) as the mobile phase, hence, the acetate ion is converted into acetic acid. The retention of acetic acid in the column, enables us to integrate the area of the peak at a given retention time to obtain the acetic acid concentration and thereby the NaOAc concentration.

A typical experiment to obtain a single solubility point (i.e., one concentration estimate at a given temperature and at a given solvent-antisolvent mixture composition) can take around two days, which makes the whole procedure rather cumbersome and tedious; such experimental complexity is unavoidable considering the absolute need to avoid exposure of the samples to ambient moisture.

\subsection{Simulations}\label{sec:simulations}

As discussed in the introduction, our simulation was made possible by the use of the WTMetaD\cite{Barducci2008} enhanced sampling method that extends the time scale limit and the use of the constant chemical potential method\cite{Perego2015} that much reduces finite-size effects.  Details on the setup of the simulations can be found in the  Supporting Information (SI) in Sections \ref{sec:S2}, \ref{sec:S3}, and \ref{sec:S4}.

With these methods we study the growth and dissolution of NaOAc ions at the  kink sites ({\it kink} in short) of a crystal surface.
After a new  kink is grown, the surface free energy does not change since the addition of a new  kink only moves the kink by one step but otherwise the kink structure remains chemically identical.
For this reason, kink growth and kink dissolution enable us to extract the free energy difference between the state of a growth unit dissolved in solution and that of a growth unit integrated within the crystal\cite{Chernov1998}.
However, NaOAc is comprised of two growth units, not just one as organic crystals. For NaOAc, both Na$^+$ and AcO$^-$ need to attach and detach during the simulation of growth and dissolution to enable the estimation of solubility. Two kink growth events, one for each ion, have to  take place to return  to the initial surface free energy. 

We expose the $\{200\}$ face of anhydrous NaOAc polymorph I\cite{Hsu1983} (see SI Section \ref{sec:S2}) to the solution, since it is the only monoatomic face of the polymorph, which has a zero electrostatic dipole moment perpendicular to the face, and therefore it is also the only stable monoatomic face. Ionic crystal faces with a non-zero dipole moment perpendicular to the surface are not stable \cite{Tasker1979}.
We consider the kinks along the edge in crystal lattice direction $[010]$, since we observed in unbiased MD simulations, that this edge is the most stable one on this specific face. The Na$^+$ and AcO$^-$ ions alternate in the row along the $[010]$ direction (see also SI Section \ref{sec:S2}).

\begin{figure}[!htbp]
\begin{center}
\includegraphics[width=14cm]{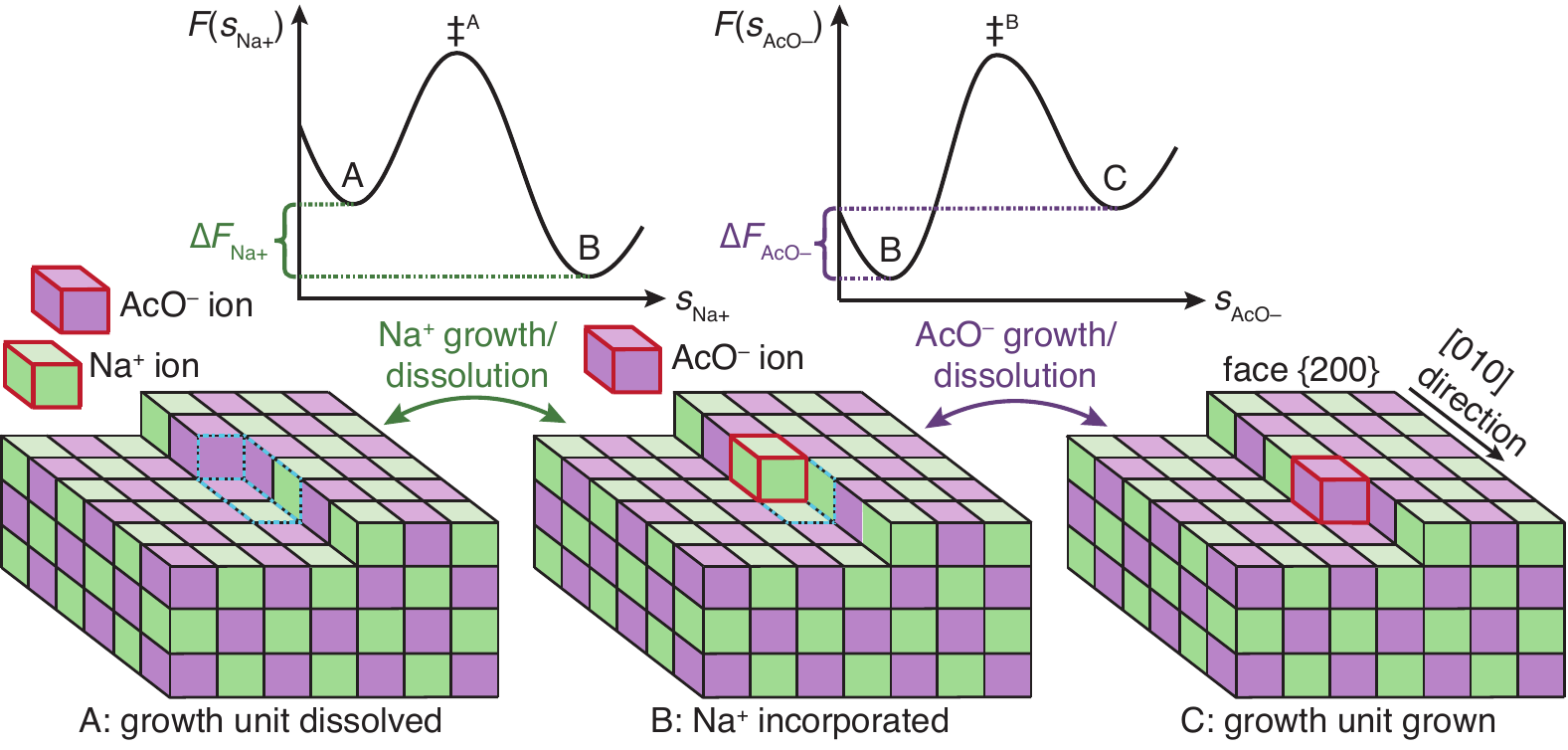}
\end{center}
\caption{Schematic of the kink growth and dissolution process of an ion dimer for NaOAc polymorph I at surface $\{200\}$ along edge [010]. Na$^+$ and AcO$^-$ growth units are shown as green and purple cubes.
State A shows a dissociated Na$^+$ and AcO$^-$ (red frames) in solution and a crystal surface with a kink site; the dimeric unit, which is about to grow, is framed with dashed blue lines.
In state B, the Na$^+$ has grown at its kink site, while the AcO$^-$ ion's kink site is still dissolved.
In state C, both ions are incorporated into the fully crystalline dimeric unit.
The growth process takes place as a sequence A $\rightarrow$ B $\rightarrow$ C, and the reverse sequence occurs for dissolution. Over each double sided arrow in the Figure, indicating the growth and dissolution process of the ions, the corresponding schemes of the free energy surfaces, $F$, are shown. The $F$ are functions of the crystallinity CVs of Na$^+$ and AcO$^-$, i.e. $s_{\text{Na}+}$ and $s_{\text{AcO}-}$, which describe the states A and B, as well as B and C. The transition states, $\ddagger^{A}$ and $\ddagger^\text{B}$, indicate the ion's growth and dissolution are activated processes.
Due to the sequential growth and dissolution, the $F$ are sampled in separate simulations for each ion at a given mole fraction, $\chi$.
The free energy difference of the dimeric unit is the sum $\Delta F = \Delta F_{\text{Na}+} + \Delta F_{\text{AcO}-}$.
$\chi$, where $\Delta F = 0$, equates to the solubility.} \label{fig:scheme}
\end{figure}

Unbiased MD simulations indicate that at concentrations close to the solubility limit, the molecules in solution are fully dissociated and enter the kink sites individually, not as a dimer.
We infer that in such a scenario the attachment of Na$^+$ and AcO$^-$ as a pair to the specific kink site is less likely since it requires a very rare process of simultaneous desolvation of both the ions and the kink sites, followed by their crystallization as a dimer.
At concentrations around solubility, the dimeric unit grows invariably according to the sequence of events illustrated in Figure \ref{fig:scheme}. We shall call the site that incorporates the kink sites of both ions the  dimeric unit\cite{Dogan2016}. 

In state A, the site of the dimeric unit is fully solvated. For the crystal to grow, first a Na$^+$ ion enters its kink site position leading to state B, which has one Na$^+$ ion incorporated into the dimeric unit but no AcO$^-$ ion. The scheme for the free energy surface (FES) of the Na$^+$ growth is shown in Figure \ref{fig:scheme} above the illustrations of state A and B. The quantity $s_{\text{Na}+}$ is a CV, which defines the solvated and crystalline Na$^+$ kink site, A and B, in the dimeric unit. The transition state is labeled by the symbol $\ddagger^\text{A}$, and its higher position between states A and B in the FES indicates that the Na$^+$ growth as well as its dissolution are activated processes. The energy difference for the grown and dissolved Na$^+$ kink site is given by $\Delta F_{\text{Na}+} = F_\text{B} - F_\text{A}$.

Only once the Na$^+$ ion is adsorbed, can the AcO$^-$ ion get attached to its kink site. The incorporation of the AcO$^-$ ion transforms the system from state B to state C along the crystallinity CV, $s_{\text{AcO}-}$. The free energy profile associated to this step is shown in the schematic. Similar to the Na$^+$ case, here also the AcO$^-$ attachment involves overcoming a free energy barrier (with transition state $\ddagger^\text{B}$) and thus it is an activated process. The free energy difference between the crystalline and dissolved AcO$^-$ kink site is computed as $\Delta F_{\text{AcO}-} = F_\text{C} - F_\text{B}$. During the dissolution process, the detachment of the ions follow the opposite sequence, i.e., the AcO$^-$ ion first, followed by the Na$^+$ ion.

The surface energy remains unaltered going from state A to C due to the conservation of the total number and type of ions residing at kink sites and edges on the surface. The free energy difference solely originates from the attachment of a new dimeric unit of NaOAc to the crystal, and the free energy difference between the crystallized and the dissolved states is given by, $\Delta F = \Delta F_{\text{Na}+} + \Delta F_{\text{AcO}-}$.
Therefore this particular kink growth process allows predicting solubility, $\chi_\text{sim}^*$ as the mole fraction, $\chi$, at which $\Delta F = 0$.

To overcome the time scale limitations encountered in kink growth and dissolution, we apply enhanced sampling via WTMetaD.
In the WTMetaD method, a bias potential is constructed as a function of a few selected slow degrees of freedom, i.e. the biased CVs, and once applied, during the simulation it discourages the system from visiting the already sampled states and thereby allows the system to come out of a free energy well, here in our context, the solvated- or the crystallized kink site state, in reasonable simulation times.

In the specific case of NaOAc, each ion undergoes the following steps during growth. The ion diffuses to the kink site, then the kink site and the ion undergo desolvation, so that the ion can adsorb onto the kink site.
Diffusion and desolvation are slow processes\cite{Chernov1998,Li2016b,Joswiak2018a}. To describe the diffusion and adsorption of each ion, we define the biased CV as a function\cite{Mendels2018a,Piccini2018,Bjelobrk2019,Rizzi2020} of the local densities of solute, as well as solvent and antisolvent, at the kink site. In WTMetaD simulations, the solute density part of the biased CV enhances the ion's diffusion towards- as well as its adsorption at the kink site, while the solvent-antisolvent density part enhances the desolvation of the kink site so that the ion can adsorb, and vice versa in the dissolution process.
The functional form of the biased CV is discussed in the SI (Section \ref{sec:S4.3}) and in our previous work\cite{Bjelobrk2021}.
To compute the energy differences, $\Delta F_{\text{Na}+}$ and $\Delta F_{\text{AcO}-}$, the trajectories of the biased CV for each ion have to be reweighed\cite{Tiwary2014} with a set of CVs that capture the crystallinity of the specific ion's kink site (see SI Sections \ref{sec:S4.4} and \ref{sec:S5} for details).

The depletion of sodium acetate from the bulk solution during the kink growth is prevented by the use of the C$\mu$MD algorithm\cite{Perego2015}, which keeps the solution's chemical potential constant in the proximity of the crystal surface. The algorithm was originally developed for a binary solution, consisting of a molecular solute and water, and it was later extended to a ternary system, consisting of a dissociated salt aqueous solution \cite{Karmakar2019}. In this work, we have to keep the chemical potential constant for a four-component system, consisting of Na$^+$, AcO$^-$, solvent, and antisolvent, which is achieved by properly choosing the simulation parameters (see also SI Section \ref{sec:S3}).

Since the NaOAc crystal with a kink site is in a metastable state, it is likely that during the simulation crystalline molecules of the unfinished surface layer dissolve. This would alter the biased kink site's environment and would lead to difficulties in the sampling. We prevent this undesired effect by the introduction of a harmonic potential, which prevents the ions from dissolution while at the same time not interfering with the natural vibrations of the ions in their lattice positions.

For all investigated molecules\cite{Mikhail1963}, the General AMBER force fields (GAFF) \cite{Wang2004,Wang2006,vanderSpoel2012,Gaussian09,Bayly1993,daSilva2012} were used with full atomistic description. Force field parameters of NaOAc were taken from Kashefolgheta \emph{et al.} \cite{Kashefolgheta2017,Joung2009,Jorgensen1983}. The NaOAc force fields with full point charges produce a considerably larger melting temperature\cite{Morris1994} of the NaOAc crystal compared to experiments\cite{Lide1992}. This would drastically lower the solubility and thus deviate dramatically from experiments\cite{Bjelobrk2021}.
This is not unexpected, as in reality, there should be considerable charge transfer between the two oppositely charged ion pairs. To incorporate such charge transfer effect, {\em albeit} approximately, we have scaled down the charges to 0.807 (see other studies involving ions in References \citenum{Leontyev2010,Doherty2017,Zeron2019}).  Now the Na$^+$ and AcO$^-$ force fields have charges +0.807 and -0.807, respectively, and reproduce the experimental melting temperature. We provide further discussion on the impact of charges on the estimated melting point and solubility in Sections \ref{sec:S1.2} and \ref{sec:S7} of the SI.

\begin{figure}[!htbp]
\begin{center}
\includegraphics[width=5cm]{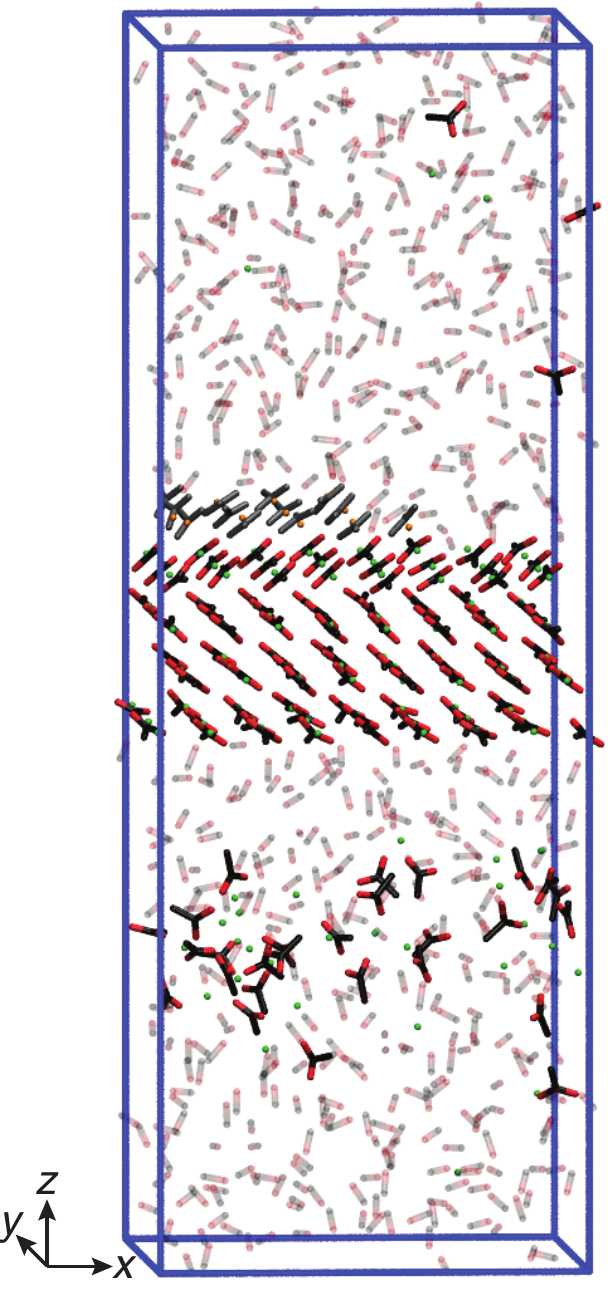}
\end{center}
\caption{Visualization\cite{Humphrey1996} of the kink growth simulation setup for Na$^+$ grown in pure MeOH solution. The biased kink site is positioned in the center of the upper surface layer. AcO$^-$ and Na$^+$ of the unfinished layer are colored in grey and orange respectively. Atoms of all other ions and molecules are colored in black for carbon, red for oxygen, and green for sodium. MeOH are shown in faded colors and hydrogens are omitted for clarity.} \label{fig:visualization}
\end{figure}

A representative visualization of the simulation setup is shown in Figure \ref{fig:visualization} for the case of Na$^+$ grown in pure MeOH solution at a solute mole fraction of $\chi = 0.0138$. For all simulations, face $\{200\}$ of anhydrous NaOAc polymorph I\cite{Hsu1983} was exposed to the solution. The unfinished surface layer was cut along the $[010]$ direction. The growth units along this specific edge are comprised of one Na$^+$ and one AcO$^-$. As previously discussed, growth along these edges consists of the integration of the Na$^+$ ion first and then of the AcO$^-$ ion, and vice versa for dissolution. This decoupled growth allows simulating the growth and dissolution of each ion separately. For the AcO$^-$ growth and dissolution simulations, the Na$^+$ ion of the biased dimeric unit was considered as part of the unfinished surface layer.

All simulations were performed with Gromacs 2016.5\cite{Berendsen1995, Lindahl2001, Spoel2005, Hess2008b, Abraham2015} patched with a custom version of Plumed 2.5.0\cite{Tribello2014}. Simulations were run at a time integration step of 0.002 ps, whereas the covalent bonds involving hydrogen atoms were constrained with the LINCS algorithm\cite{Hess2008b, Hess2008a}.
For long-range electrostatics, the particle mesh Ewald algorithm\cite{Ewald1921,Darden1993} was used, and the non-bonded interactions (electrostatic and van der Waals) cutoff was set to 1 nm. 
The simulations were run at a temperature of $T = 298.15$ K using the stochastic velocity rescale thermostat \cite{Bussi2009} in the NVT ensemble. The simulation box lengths were fixed at their average values which were obtained from NPT simulations run at 1 bar pressure using the Parrinello-Rahman barostat\cite{Parrinello1981}. A simulation time of at least 1.0 $\mu$s was necessary to obtain sufficient growth and dissolution events for each ion and to obtain a converged $\Delta F$.

\section{Results}

We have measured through experiments and estimated through MD simulations, the solubility of sodium acetate in six different solvent-antisolvent mixtures, using the methodologies described in Sections \ref{sec:experiments} and \ref{sec:simulations}, respectively.

\begin{figure}[!htbp]
\begin{center}
\includegraphics[width=\textwidth]{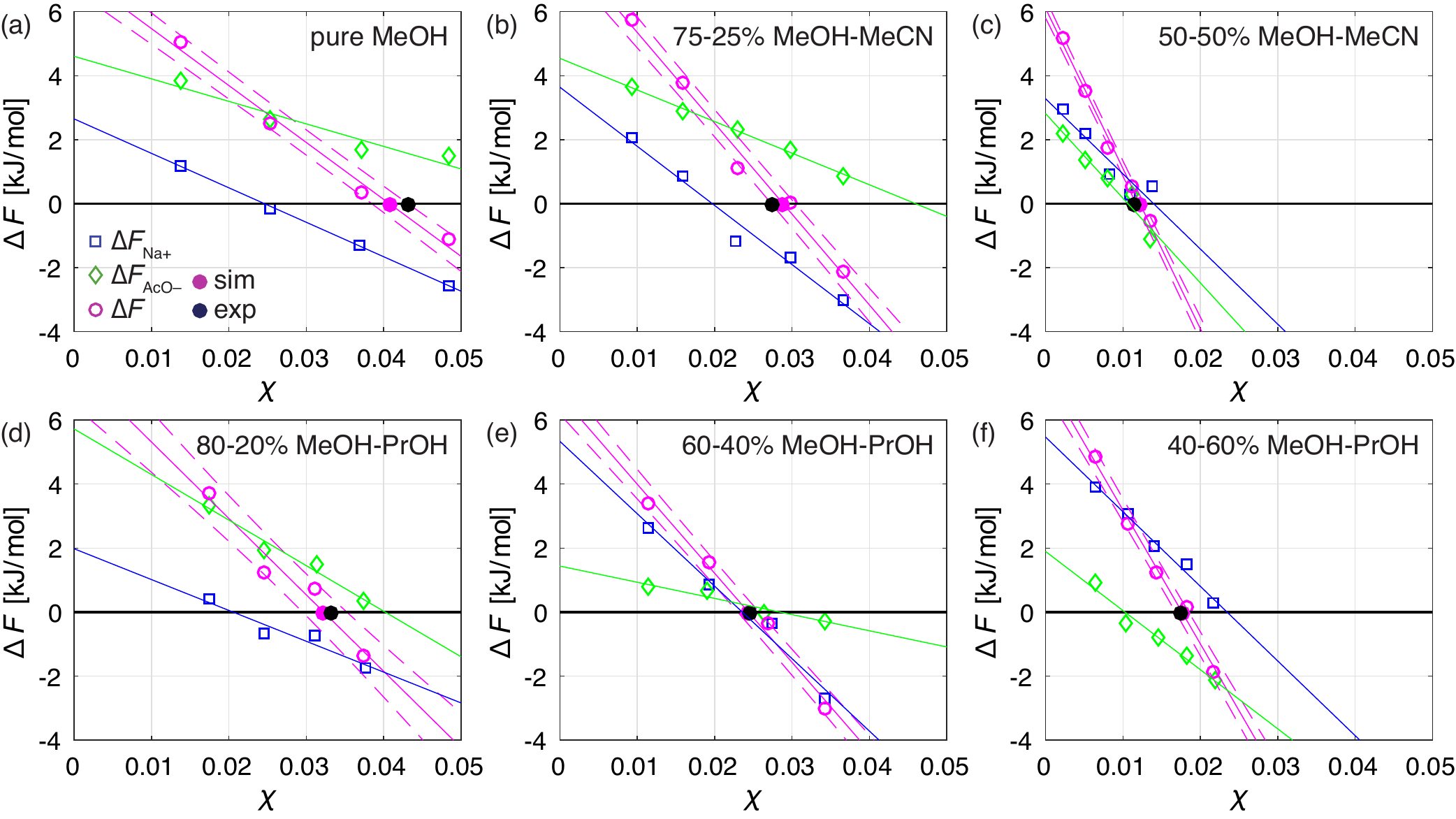}
\end{center}
\caption{Sampled energy differences between grown and solvated kink sites of Na$^+$, $\Delta F_{\text{Na}+}$ (blue boxes), and AcO$^-$, $\Delta F_{\text{AcO}-}$ (green diamonds), as well as the dimeric unit, $\Delta F = \Delta F_{\text{Na}+} + \Delta F_{\text{AcO}-}$ (purple circles), as a function of NaOAc mole fraction, $\chi$, and compared to the experimental solubility values (black filled circles). Straight purple lines represent the linear regression of $\Delta F$ and the dashed lines are the corresponding standard deviations. The blue and green straight lines are the linear regressions of $\Delta F_{\text{Na}+}$ and $\Delta F_{\text{AcO}-}$ respectively.
The predicted solubilities (purple filled circles) correspond to the mole fraction at $\Delta F = 0$ obtained through a linear regression of $\Delta F$ as a function of $\chi$.
The results are shown for the systems of crystalline NaOAc exposed to following solutions:
a) pure MeOH;
b) \num{75}-\num{25}\si{\percent} MeOH-MeCN;
c) \num{50}-\num{50}\si{\percent} MeOH-MeCN;
d) \num{80}-\num{20}\si{\percent} MeOH-PrOH;
e) \num{60}-\num{40}\si{\percent} MeOH-PrOH;
f) \num{40}-\num{60}\si{\percent} MeOH-PrOH.
It is important to note that some of the presented $\Delta F_{\text{Na}+}$ and $\Delta F_{\text{AcO}-}$ points are the averages of simulation repetitions (see SI Section \ref{sec:S6}).
} \label{fig:results_ms}
\end{figure}

Figure \ref{fig:results_ms} shows the results for all the solution compositions investigated:
pure MeOH (panel (a)), 
\num{75}-\num{25}\si{\percent} MeOH-MeCN (panel (b)),  
\num{50}-\num{50}\si{\percent} MeOH-MeCN (panel (c)), 
\num{80}-\num{20}\si{\percent} MeOH-PrOH (panel (d)), 
\num{60}-\num{40}\si{\percent} MeOH-PrOH (panel (e)), and
\num{40}-\num{60}\si{\percent} MeOH-PrOH (panel (f)).
In this figure, $\Delta F_{\text{Na}+}$ corresponds to blue boxes, $\Delta F_{\text{AcO}-}$ to green diamonds, 
and $\Delta F$ to purple circles. The blue and green lines are linear regressions of $\Delta F_{\text{Na}+}$ and $\Delta F_{\text{AcO}-}$ as functions of $\chi$ respectively.
We obtain $\chi_\text{sim}^*$ through linear regression of $\Delta F$ as a function of $\chi$, plotted as a purple line, with the corresponding lower and upper bounds of the standard deviation plotted as dashed purple lines (which contain the 68 \% confidence interval). The estimated solubility, $\chi_\text{sim}^*$, is the intersection point of the regression line with the horizontal axis (the corresponding point is the purple filled circle). The experimental solubilities, $\chi_\text{exp}^*$, obtained using the methodology presented in Section \ref{sec:experiments}, are plotted as black filled circles. The values obtained through experiments and MD simulations are also listed in Table \ref{tab:results_ms}.

There are a few remarks worth making.

First, in all cases linear regression of the predicted values of $\Delta F_{\text{Na}+}$, $\Delta F_{\text{AcO}-}$, and of their sum, i.e., $\Delta F$, exhibits a satisfactory goodness-of-fit. Therefore, the quantitative and qualitative behavior of the corresponding trend lines is worth analysing.

Second, when considering the intersection of the free energy change upon growth of one dimeric unit at the kink site, i.e., $\Delta F=0$ at the solute solubility concentration, $\chi^*$, one observes that as expected solubility is the highest in pure MeOH, and it decreases with increasing levels of antisolvent in the solution, in the case of both propan-1-ol and acetonitrile.

\begin{table}[!htbp]
\centering
\captionsetup{justification=centering}
\caption{Values of the experimental solubilities, $\chi_\text{exp}^*$, and the respective predicted solubilities, $\chi_\text{sim}^*$, with the standard errors in parenthesis.}
\centering \label{tab:results_ms}
\begin{tabular*}{\textwidth}{@{\extracolsep{\fill}}llllllllll}
\toprule
                        & & pure MeOH  & & \multicolumn{2}{c}{MeOH-MeCN}  & & \multicolumn{3}{c}{MeOH-PrOH}       \\
   & & 100\si{\percent} & &  75-25\si{\percent} &  50-50\si{\percent} & & 80-20\si{\percent} & 60-40\si{\percent} & 40-60\si{\percent}     \\
\midrule
$\chi_\text{exp}^*$ [-] & & 0.0440     & & 0.0275        & 0.0113         & & 0.0333     & 0.0246     & 0.0175    \\
$\chi_\text{sim}^*$ [-] & & 0.0407(24) & & 0.0288(16)    & 0.0122(5)      & & 0.0321(26) & 0.0244(14) & 0.0176(9) \\
\bottomrule
\end{tabular*}
\end{table}

Third, when comparing the trend lines exhibited by each individual free energy difference
in the six different cases illustrated in Figure \ref{fig:results_ms}, the following observations can be made: (i) the $\Delta F$ trend lines intersect the horizontal axis from right to left obviously in the same sequence as the decreasing values of solubility, $\chi^*$, the steeper the line the lower the solubility; (ii) the $\Delta F_{\text{Na}+}$ are much less affected by the change in solvent system, with their intersection with the horizontal line $\Delta F_{\text{Na}+}=0$ between $\chi \approx 0.015$ and $\chi \approx 0.025$; the $\Delta F_{\text{AcO}-}$ are strongly affected by the presence of the antisolvent, with their intersection with the horizontal line $\Delta F_{\text{AcO}-}=0$ between $\chi \approx 0.01$ and $\chi \approx 0.065$ (the high values being those of the systems with MeOH only or with a lot of it, and the low values being associated to the systems with a lot of antisolvent). 

Fourth, the last observation can be given a mechanistic interpretation, which is supported by unbiased simulations not reported here for brevity. The adsorption and inclusion of the acetate ion, AcO$^-$, onto the kink site is strongly affected by the local environment, which is obviously dependent on the solution composition and on the concentration of the antisolvent. In this study the two antisolvents are bulkier and less polar than methanol, the solvent. As a consequence, the latter tends to occupy the kink site with stronger bonds than the former, thus making it more difficult for the acetate ion to adsorb and to be incorporated in the crystal lattice. The ensuing lower energy difference $\Delta F_{\text{AcO}-}$ in the case of higher antisolvent concentrations leads to lower solubility. This is indeed the case, since, as we have observed, adsorption and inclusion of the sodium ion, Na$^+$, onto the kink site is rather weakly affected by the presence of the antisolvent and by its concentration, which is possibly due to its tiny size as compared to the acetate ion and to the solvent and antisolvent molecules.

The fifth observation is that - as it is apparent, both in the table and in the figure - the predicted and the measured solubility values are in excellent agreement, the maximum relative difference being \SI{8}{\percent} for the case \num{50}-\num{50}\si{\percent} MeOH-MeCN. The values of solubility predicted by the MD simulations and those measured experimentally match so well despite they have been obtained using two completely different methods, each of which representing a significant novelty in modeling and in experimental characterization.

Finally, it is worth noting that thanks to the NaOAc force fields\cite{Kashefolgheta2017} used for the simulations, with proper charge scaling, we are able to accurately predict the experimental melting temperature and thereby the experimental solubility under all conditions explored in this work. This is consistent with our previous observations on estimating the solubility of organic molecules using MD simulations\cite{Bjelobrk2021}.

\section{Conclusions}

In this work, we present a molecular dynamics simulation method for the prediction of the solubility of organic salts and validate the predicted solubilities with the experiment outcome. The simulation method focuses on the growth and dissolution of ions at kink sites and identifies the solubility, where the energy difference between the ion-pair crystallized at the kink sites and dissolved in solution equals zero.
We showcase the method's potential by applying it to estimate the solubility of anhydrous sodium acetate in a variety of solvent-antisolvent mixtures.
In parallel, we measure the salt's solubility with an in-house experimental setup, which, coping with the difficulties due to the salt's strong hygroscopicity, allows to estimate the solubility under moisture-free conditions. We obtain excellent agreement between experiments and simulations. 

The presented simulation setup is general and can be used to predict solubilities of organic salts in complex growth environments. The method becomes particularly interesting for growth environments, which are very difficult to attain and control experimentally, such as for substances, which are sensitive to air humidity, as in the studied case of anhydrous sodium acetate.
Computing solubility using our method can allow significant acceleration and cost-reduction in the estimation of solubility in compounds of interest under varying conditions. These advantages become especially compelling when high throughput endeavours are considered, in particular if non-standard experimental equipment, such as the one presented here, needs to be utilized. Moreover, with the rapid improvement in accuracy and efficiency of MD simulations due to utilization of GPUs and machine learning based force fields \cite{Behler2007,Zhang2018}, the advantages of using the approach introduced here is bound to increase further substantially.

An interesting application of the simulation method can be in the field of counter ion screening. Counter ions of organic salt APIs are often screened to tailor the substance's solubility\cite{Paulekuhn2007}. 
Through simulations, a much clearer picture can be gained on how the particular counter ion behaves in the solution and at the kink sites. This in turn can guide and speed up the counter ion selection procedure to attain the desired solubility properties of the specific substance in the solution of interest.

\section*{Acknowledgements}
Z. B. and M. M. are thankful to Novartis Pharma AG for their partial financial support to this project. Z. B., A. K. R., and M. M. thank Bianca Popa, Ayoung Song, Johann Bartenstein, and Luca Bosetti for performing parts of the experiments.
Z. B. thanks Riccardo Capelli for providing the force field for propan-1-ol and Pablo Piaggi, Ruben W\"alchli, Thilo Weber, Philipp M\"uller, and Marco Holzer for valuable discussions.
The computational resources were provided by ETH Z\"urich and the Swiss Center for Scientific Computing at the Euler Cluster.

\bibliographystyle{unsrt} 

\bibliography{bibliography} 

\newpage
\begin{center}
\vspace{5mm}
\LARGE{\textsc{Supporting information}}
\vspace{2mm}
\end{center}

\setcounter{section}{0}
\renewcommand*{\theHsection}{chX.\the\value{section}}

\setcounter{figure}{0}
\setcounter{table}{0}

\makeatletter
\@addtoreset{section}{part}
\renewcommand \thesection{S\@arabic\c@section}
\renewcommand \thetable{S\@arabic\c@table}
\renewcommand \thefigure{S\@arabic\c@figure}
\renewcommand \theequation{S\@arabic\c@equation}
\makeatother


\section{Force fields}\label{sec:S1}

In this work we have used the general AMBER force fields (GAFF)\cite{Wang2004, Wang2006} for all our solvent and antisolvent molecules. The force fields have full atomistic description and the bonds involving hydrogens were fixed at their equilibrium value. Force field parameters for MeOH and MeCN were taken from van der Spoel \emph{et al.} \cite{vanderSpoel2012} and the NaOAc force fields were taken from Kashefolgheta \emph{et al.} \cite{Kashefolgheta2017}. The propan-1-ol and sodium acetate force fields are discussed in the following two Sections \ref{sec:S1.1} and \ref{sec:S1.2}.

\subsection{Propan-1-ol force field} \label{sec:S1.1}  

\begin{figure}[!htbp]
\begin{center}
\includegraphics[width=\textwidth]{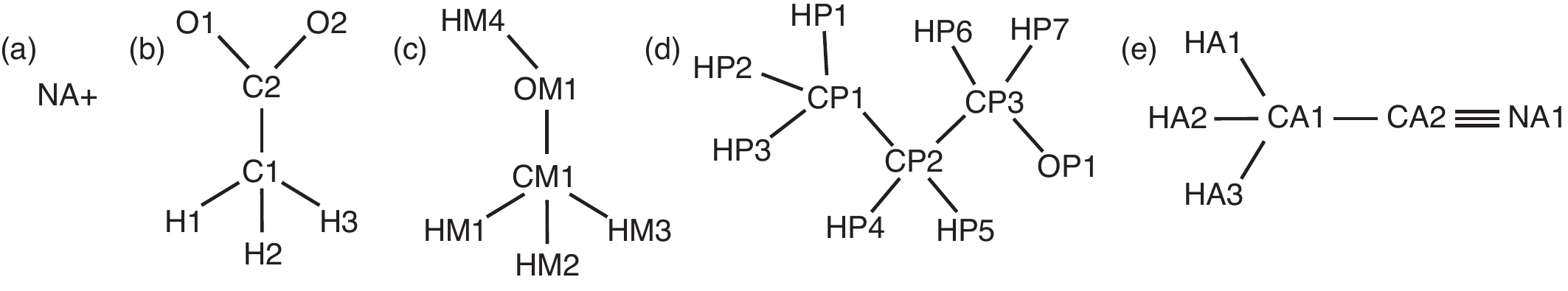}
\end{center}
\caption{Structural formulas with the used atom names for: a) sodium ion, b) acetate ion, c) methanol, d) propan-1-ol, and e) acetonitrile.} \label{fig:labels}
\end{figure}

The electrostatic potential of propan-1-ol was calculated with Gaussian 09 \cite{Gaussian09} at the B3LYP/6-31G(d,p) level and the partial charges were fitted to the restrained electrostatic potential charges \cite{Bayly1993, daSilva2012}.
The obtained GAFF parameters of propan-1-ol are listed in Table \ref{tab:propan-1-ol} and
the corresponding structural formula with the atom names used in Table \ref{tab:propan-1-ol} is shown in Figure \ref{fig:labels}e.

To roughly estimate the force field's ability to reproduce reality, we compare the density of a simulated propan-1-ol liquid with the experimental density of $\rho_\text{PrOH}^\text{exp}$ = 799.5 kg/m$^3$ at ambient conditions \cite{Mikhail1963}. We have therefore performed a simulation run
of a box containing 350 liquefied propan-1-ol molecules using Gromacs 2016.5. The simulation was performed at $NPT$ conditions with the Parrinello-Rahman barostat \cite{Parrinello1981} using isotropic pressure coupling and the velocity rescale thermostat \cite{Bussi2009} at $p$ = 1 bar and $T$ = 298.15 K. The simulation was run with periodic boundary conditions. The simulated liquid has an average density of $\rho_\text{PrOH}^\text{sim}$ = 789.2 kg/m$^3$, which was averaged over 25 ns, and is in reasonable accordance with experiment (-1.29 \% difference).

\begin{table}[!htbp]
\caption{GAFF parameters for propan-1-ol.}
\centering
\begin{tabular*}{\textwidth}{@{\extracolsep{\fill}}ccccccc}
\toprule
atom	& GAFF		& RESP	     & mass	   & $x$	 & $y$ 	   & $z$	 \\ 
name	& atom type	& charge [e] & [g/mol] & [nm]	 & [nm]	   & [nm]   \\
\midrule
CP1     & c3        & -0.265718  & 12.010  & -0.178  &  0.037  &  0.121  \\
HP1     & hc        &  0.054899  &  1.008  & -0.214  & -0.062  &  0.148  \\
HP2     & hc        &  0.054899  &  1.008  & -0.263  &  0.096  &  0.086  \\
HP3     & hc        &  0.054899  &  1.008  & -0.137  &  0.084  &  0.210  \\
CP2     & c3        &  0.164627  & 12.010  & -0.070  &  0.026  &  0.011  \\
HP4     & hc        & -0.003642  &  1.008  & -0.113  & -0.024  & -0.076  \\
HP5     & hc        & -0.003642  &  1.008  &  0.012  & -0.035  &  0.047  \\
CP3     & c3        &  0.361627  & 12.010  & -0.018  &  0.165  & -0.029  \\
HP6     & h1        & -0.048986  &  1.008  & -0.100  &  0.226  & -0.067  \\
HP7     & h1        & -0.048986  &  1.008  &  0.026  &  0.215  &  0.058  \\
OP1     & oh        & -0.750180  & 16.000  &  0.081  &  0.151  & -0.131  \\
HP8     & ho        &  0.430202  &  1.008  &  0.113  &  0.239  & -0.155  \\
\bottomrule
\end{tabular*}
\label{tab:propan-1-ol}
\end{table}

\subsection{Sodium acetate force fields} \label{sec:S1.2} 

For the NaOAc ion pair, force fields reported by Kashefolgheta \emph{et al.} \cite{Kashefolgheta2017} were used, which are comprised of GAFF compatible parameters. The force field of AcO$^-$ is optimized with the TIP3P water model \cite{Jorgensen1983} and Na$^+$ model of Joung \emph{et al.} \cite{Joung2009} to reproduce experimental hydration- and solvation free energies at low solute concentrations (0.5 M). Unoptimized GAFF parameters for NaOAc would overestimate the ion-ion and ion-solvent interactions. The interested reader is referred to Reference \citenum{Kashefolgheta2017} for further details.

As we are dealing here with the crystalline NaOAc phase as well, we require a force field which is capable of reproducing the experimental melting temperature of NaOAc, and not primarily solvation free energies in water. Since the reported NaOAc force fields overestimate the experimental melting temperature, $T_\text{m}^\text{exp}$, considerably by over 300 K, we need to perform a further fix-up of the force fields for our purposes of obtaining reliable solubility estimates for NaOAc. Without scaling, the solubility predictions will provide results at least an order of magnitude smaller than experiments (see also Section \ref{sec:S7}). We therefore linearly scale the partial charges of the NaOAc atoms by a scaling factor, $0 < q < 1$, to reduce the simulated NaOAc melting temperature, $T_\text{m}^\text{sim}$, to meet the experimental one, $T_\text{m}^\text{exp}$ = 597 K \cite{Lide1992}. Charge scaling is commonly used to improve the performance of coulombic interactions of non-polarizable force fields in condensed matter simulations and to make them more realistic \cite{Leontyev2010,Doherty2017}.

\begin{figure}[!htbp]
\begin{center}
\includegraphics[width=\textwidth]{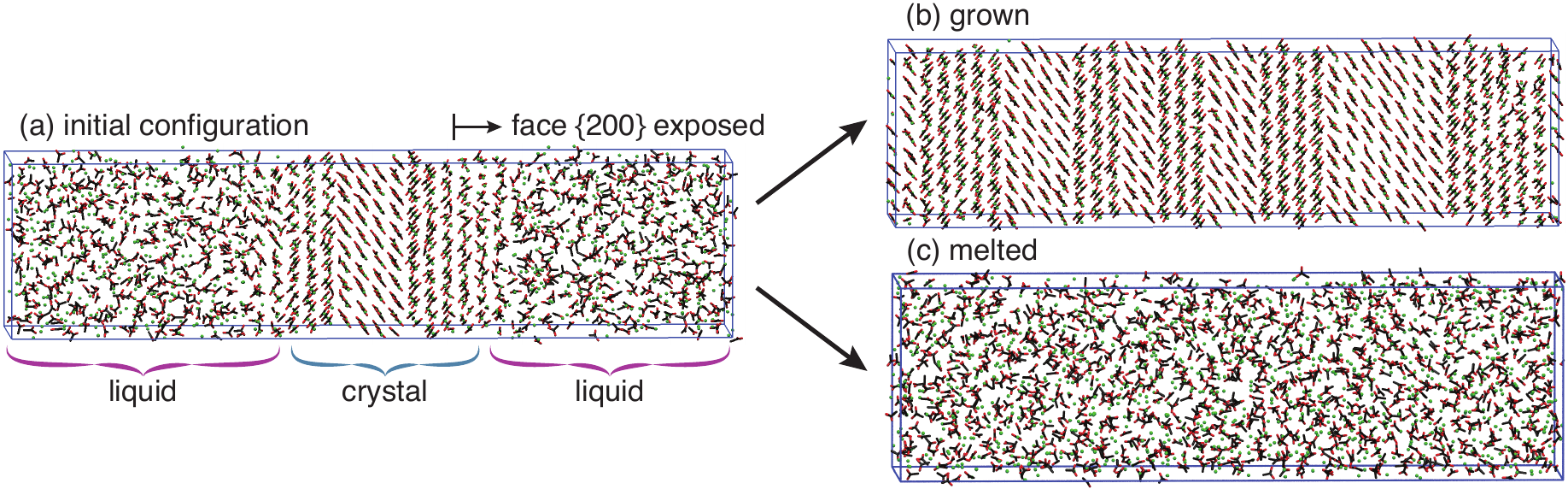}
\end{center}
\caption{Visualizations of NaOAc crystal-melt coexistence simulations. a) Initial configuration with a crystalline phase (in the center of the box) submerged in the melted phase. b) Fully grown crystalline phase. c) Fully dissolved crystalline phase.} \label{fig:melt_simulations}
\end{figure}

To obtain the melting temperature of the NaOAc force fields for a given $q$ we use crystal-melt coexistence MD simulations \cite{Morris1994}. 
Figure \ref{fig:melt_simulations}a shows a representative visualization of the simulation box at the beginning of the simulation, where a crystal of the anhydrous sodium acetate polymorph I is exposed to a sodium acetate melt using periodic boundary conditions. The initial crystal has roughly the dimensions of 3 $\times$ 2.5 $\times$ 3 nm$^3$ and occupies a third of the simulation box volume, while the other two thirds are occupied by the melt.
The crystal face $\{200\}$ was exposed to the melt since it is a fast growing face which allows us to observe growth or dissolution of the crystal within simulation times of 50 ns. Figures \ref{fig:melt_simulations}b and \ref{fig:melt_simulations}c show visualizations of the final configuration of the fully grown and fully dissolved crystalline phase respectively.
The simulations were performed under $NPT$ conditions with the velocity rescale thermostat \cite{Bussi2009} and the
Parrinello-Rahman barostat \cite{Parrinello1981}  at $p$ = 1 bar using semi-isotropic pressure coupling. Only the box length $L_\text{z}$ perpendicular to the crystal surface was allowed to change, while box lengths $L_\text{x}$ and $L_\text{y}$ parallel to the crystal surface were fixed at equilibrium distances.

For each $q$, sets of individual simulations at different $T$ were performed and the results are listed in Table \ref{tab:melt_results}: the value $q$ = 0.807 was identified to produce a force fields melting point of $T_\text{m}^\text{sim}$ = 596 K, which is in good agreement with its experimental counterpart. The scaled charges of the NaOAc force fields \cite{Kashefolgheta2017} are listed in Table \ref{tab:scaledcharges} and the corresponding structural formulas with the atom names are shown in Figures \ref{fig:labels}a and \ref{fig:labels}b.

\begin{table}[h]
\centering
\caption{Individual coexistence NaOAc crystal-melt simulation outcomes for different charge scaling factors $q$ and simulation temperatures $T$. d indicates full dissolution of the crystalline phase and g indicates full growth of the crystalline phase. Out of the tested values, $q$ = 0.807 provides the best agreement with the experimental melting point of $T_\text{m}^\text{exp}$ = 597 K \cite{Lide1992}.}
\begin{tabular*}{\textwidth}{@{\extracolsep{\fill}}ccccccccccccc}
\toprule
$q$   & 595 K & 596 K & 597 K & 598 K & 599 K & 600 K  & 605 K & 630 K & 635 K & 640 K & $T_\text{m}^\text{sim}$  \\
\midrule
0.805 & d     &       &       &       &        & d     &       &       &       &       & < 595 K   \\
0.806 & d     &       &       &       &        &       &       &       &       &       & < 595 K   \\
0.807 & g     & d     & d     &       &        & d     &       &       &       &       & 596 K     \\
0.808 & g     &       & g     & g     & d      & d     & d     &       &       &       & 599 K     \\
0.809 & g     & g     & g     &       &        & g     & d     &       &       &       & 601-605 K \\
0.810 & g     &       &       &       &        & g     & d     &       &       &       & 601-605 K \\
0.840 &       &       &       &       &        & g     &       & g     & g     & d     & 636-640 K \\
\bottomrule
\end{tabular*}
\label{tab:melt_results}
\end{table}

\begin{table}[!htbp]
\caption{Scaled charges used for the NaOAc force fields\cite{Kashefolgheta2017}.}
\centering
\begin{tabular*}{\textwidth}{@{\extracolsep{\fill}}ccc}
\toprule
atom  & AMBER 	   & scaled     \\ 
name  & atom type  & charge [e] \\
\midrule
C1    & CT         & -0.171567  \\
H1    & HC         &  0.002887  \\
H2    & HC         &  0.002887  \\
H3    & HC         &  0.002887  \\
C2    & C          &  0.712154  \\
O1    & OACE       & -0.678124  \\
O2    & OACE       & -0.678124  \\
NA+   & NA+        &  0.807000  \\
\bottomrule
\end{tabular*}
\label{tab:scaledcharges}
\end{table}

The NaOAc force fields scaled with $q = 0.807$ were tested whether they can reproduce the experimental crystal structure. A simulation of the NaOAc polymorph I was therefore performed and compared to the experimental crystallographic data \cite{Hsu1983}. Anhydrous NaOAc polymorph I is orthorombic and belongs to the $Pcca$ space group. 

\begin{figure}[!htbp]
\begin{center}
\includegraphics[width=\textwidth]{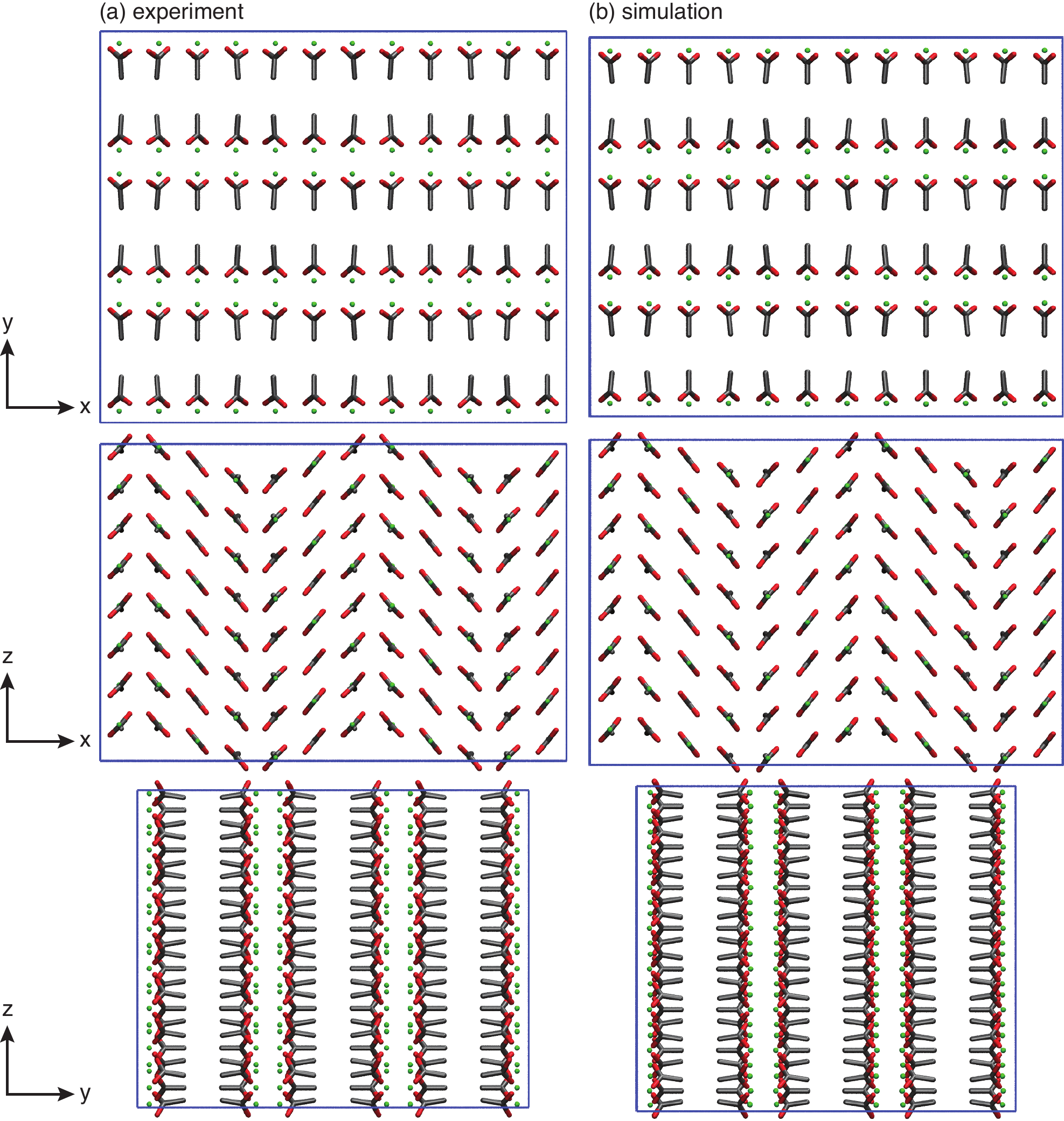}
\end{center}
\caption{NaOAc crystal visualizations projected along $z$, $y$, and $x$ directions. The green spheres denote Na$^+$ and AcO$^-$'s oxygens and carbons are colored in red and black respectively. Hydrogens are omitted for clarity. The blue lines denote the simulation box edges. a) Atom positions from the experimental XRD data. b) Averaged atom positions of the $NPT$ simulation.} \label{fig:crystalboxes}
\end{figure}

A crystal of roughly the dimensions 3.5 $\times$ 3.0 $\times$ 2.5 nm$^3$ containing 288 NaOAc ion pairs was prepared using the experimental atom positions obtained from XRD measurement\cite{Hsu1983} as initial condition. Figure \ref{fig:crystalboxes}a shows the initial condition (experimental positions of NaOAcs) of the crystal.
The simulation was then performed for 50 ns with periodic boundary conditions at $p$ = 1 bar and $T$ = 298 K using the anisotropic Parrinello-Rahman barostat \cite{Parrinello1981} and the velocity rescale thermostat \cite{Bussi2009}. 

The resulting averaged unit cell lengths, $a_\text{u}$, $b_\text{u}$, and $c_\text{u}$ and unit cell volume $V_\text{u}$ of the simulation are listed and compared to experiments in Table \ref{tab:unitcell}. Only the last 25 ns were used for the averaging. The corresponding visualization of time averaged NaOAcs positions within the crystal are shown in Figure \ref{fig:crystalboxes}b. From Table \ref{tab:unitcell} and Figure \ref{fig:crystalboxes} we can conclude that the NaOAc force fields can reasonably well reproduce the experimental polymorph values.

\begin{table}[!htbp]
\caption{Simulated vs. experimental \cite{Hsu1983} unit cell lengths, $a_\text{u}$, $b_\text{u}$, and $c_\text{u}$, and unit cell volumes, $V_\text{u}$, of sodium acetate polymorph I (orthorhombic, $Pcca$ space group).}
\centering
\begin{tabular*}{\textwidth}{@{\extracolsep{\fill}}cccc}
\toprule
              & experiment     & simulation     & difference  \\
\midrule
$a_\text{u}$  & 1.7850 nm      & 1.8136 nm      & 1.60  \%    \\
$b_\text{u}$  & 0.9982 nm      & 0.9657 nm      & -3.26 \%    \\
$c_\text{u}$  & 0.6068 nm      & 0.6219 nm      & 2.49  \%    \\
$V_\text{u}$  & 1.0812 nm$^3$  & 1.0892 nm$^3$  & 0.74  \%    \\
\bottomrule
\end{tabular*}
\label{tab:unitcell}
\end{table}

\begin{figure}[!htbp]
\begin{center}
\includegraphics[width=\textwidth]{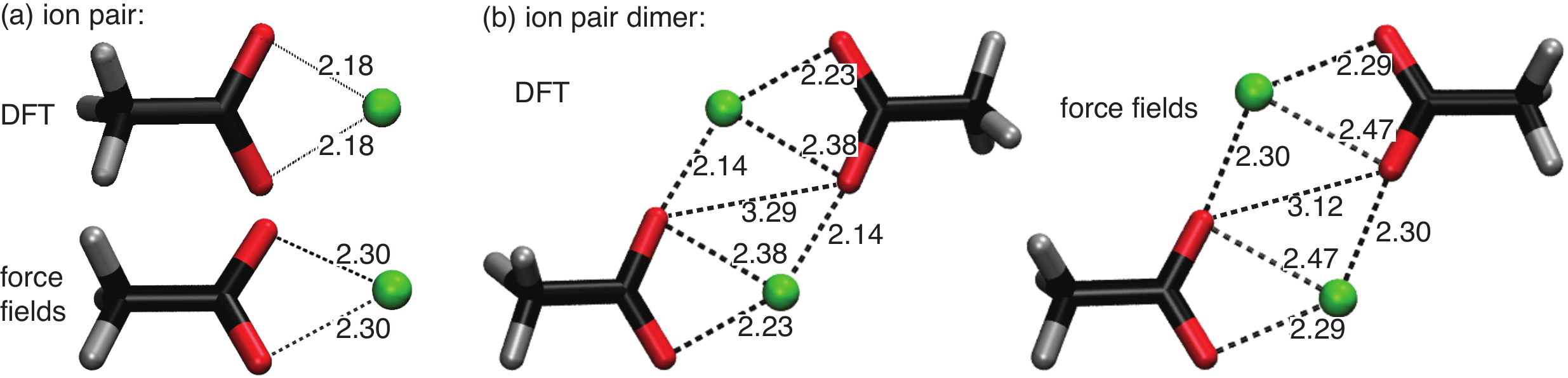}
\end{center}
\caption{Visualization of the energy minimized structures for a) NaOAc ion pair and b) NaOAc ion pair dimers in vacuum calculated with DFT (top) and the scaled NaOAc force fields (bottom). The same color scheme of the atoms as in Figure \ref{fig:crystalboxes} is used, whereat hydrogen molecules are colored in grey. Atom distances between the atoms, indicated with dashed lines, are given in \AA.} \label{fig:structureFF}
\end{figure}

However, the length $b_\text{u}$, which is along the C2-NA+ axis, is slightly underestimated in the simulations while lengths $a_\text{u}$ and $c_\text{u}$ are slightly overestimated. To understand this discrepancy between simulations and experiments we further computed the energy minimized structures for an ion pair and an ion pair dimer with the NaOAc force fields and compared them to DFT calculations. The results are visualized in Figure \ref{fig:structureFF}.
The energy minimization of the NaOAc force fields was performed with Gromacs 2016.5 \cite{Abraham2015} with the conjugate gradient algorithm with a tolerance of the maximum force of 0.001 kJ mol$^{-1}$ nm$^{-1}$.
DFT calculations were performed with Gaussian structural optimization routine at the B3LYP/6-31G(d,p) level \cite{Gaussian09}.
The AcO$^-$-Na$^+$ interaction is underestimated in the force fields as the distances between the AcO$^-$ oxygens and Na$^+$ are shorter in the DFT calculations (2.18 {\AA}) compared to the force fields (2.30 {\AA}, see Figure \ref{fig:structureFF}a). For the ion pair dimer (see Figure \ref{fig:structureFF}b), the distances between AcO$^-$ oxygens and Na$^+$ are as well shorter in the DFT calculations (2.14, 2.38, and 2.23 {\AA}) when compared to the force fields (2.30, 2.47, and 2.29 {\AA}). This coincides with observation, that the unit cell lengths $a_\text{u}$ and $c_\text{u}$ are longer in simulations compared to the experiments. Contrariwise, the AcO$^-$-AcO$^-$ distance is longer in the DFT calculation (3.29 {\AA} for the nearest oxygens) than it is for the force fields' calculation (3.12 {\AA}), which coincides with the observation of a slightly shorter unit cell length $b_text{u}$ in simulations compared to experiments.

Overall, the scaled NaOAc force fields reproduce reasonably well the crystallographic data.

\section{Simulation setup and equilibration}\label{sec:S2}

\begin{figure}[!htbp]
\begin{center}
\includegraphics[width=15.5cm]{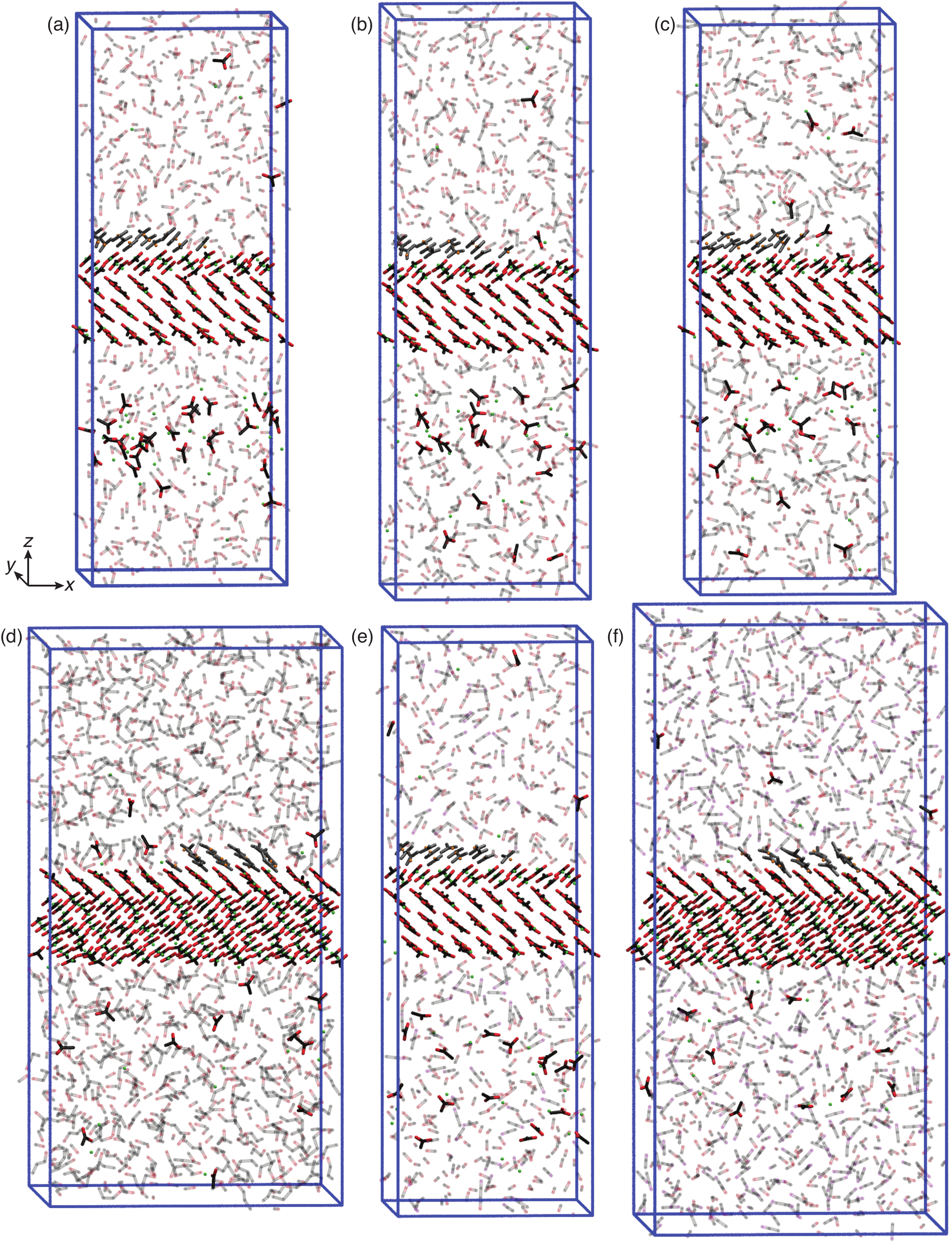}
\end{center}
\caption{Visualizations of simulation boxes used for Na$^+$ growth and dissolution simulations in each of the investigated solvent/antisolvent mixtures. The biased kink site can be found in the middle of each upper crystal surface. a) shows the case for a pure MeOH, b) 80-20\% MeOH-PrOH, c) 60-40\% MeOH-PrOH, d) 40-60\% MeOH-PrOH, e) 75-25\% MeOH-MeCN, and f) the 50-50\% MeOH-MeCN solution. The Na$^+$ and AcO$^-$ in the unfinished surface layer are colored in grey, while all other Na$^+$ ions are colored in green. Carbons are colored in black, oxygens in red, and nitrogens in purple. The solvent and antisolvent molecules are shown in faded colors. Hydrogens are omitted for clarity.
} \label{fig:setups}
\end{figure}

\begin{figure}[!htbp]
\begin{center}
\includegraphics[width=\textwidth]{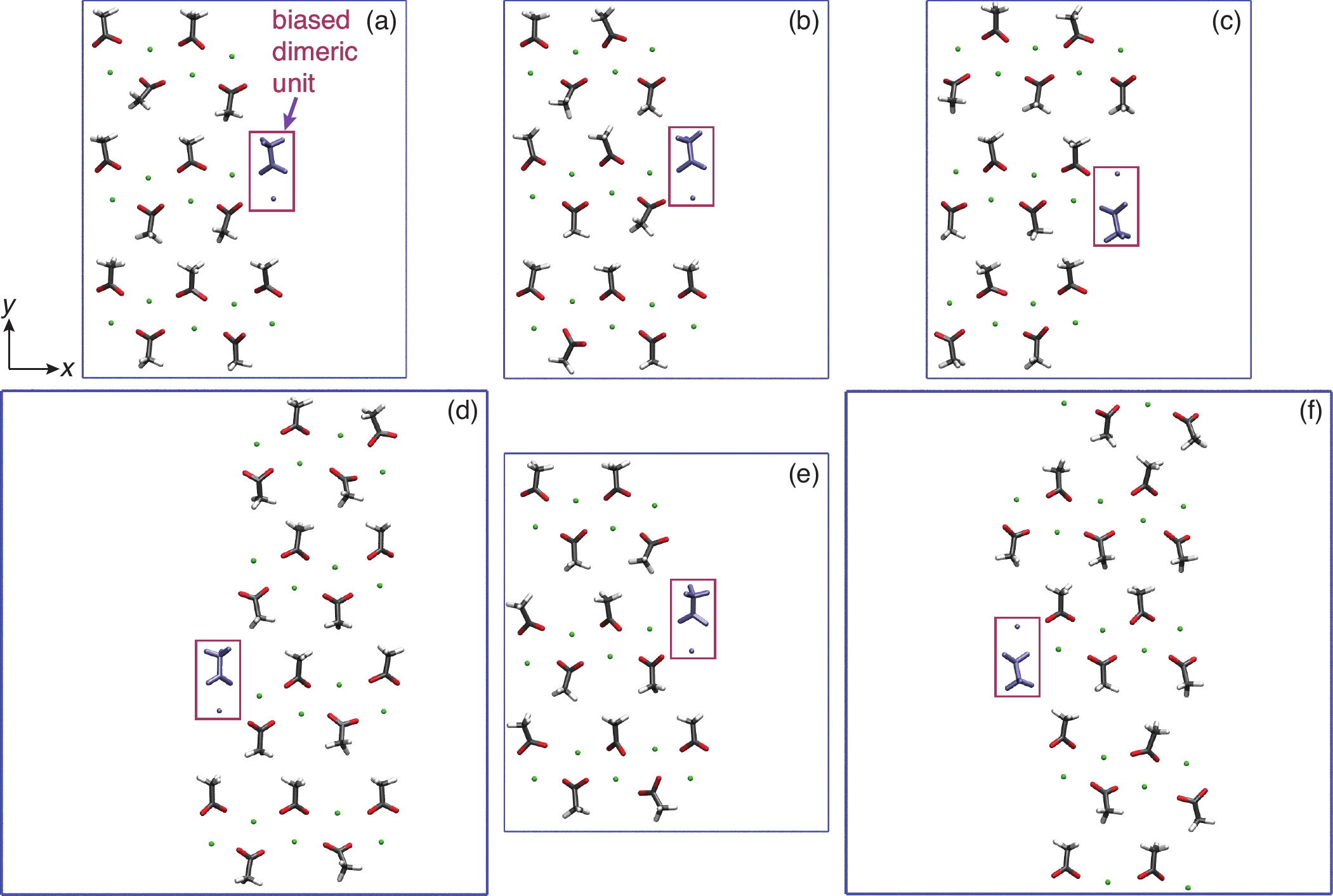}
\end{center}
\caption{Visualizations of the unfinished surface layers, seen along the $z$ axis perpendicular to the crystal surface, for a) 100\% MeOH, b) 80-20\% MeOH-PrOH, c) 60-40\% MeOH-PrOH, d) 40-60\% MeOH-PrOH, e) 75-25\% MeOH-MeCN, and f) 50-50\% MeOH-MeCN. Na$^+$ are colored in green, carbons in black, oxygens in red and hydrogens in light grey. The NaOAc dimeric unit, comprising the biased kink sites of Na$^+$ and AcO$^-$, is colored in blue and framed in red.} \label{fig:unfinishedl}
\end{figure}

In this work, we performed kink growth and dissolution sampling of Na$^+$ and AcO$^-$ at the NaOAc crystal kink site in solvent/antisolvent mixtures in accordance with the reported experiments. Namely in pure MeOH, 80-20\% MeOH-PrOH, 60-40\% MeOH-PrOH, 40-60\% MeOH-PrOH, 75-25\% MeOH-MeCN, and 50-50\% MeOH-MeCN.
The simulation box specifications for the six systems are listed in Table \ref{tab:boxspecs}. At high antisolvent concentrations, i.e. 40-60\% MeOH-PrOH and 50-50\% MeOH-MeCN, larger simulation boxes were used in order to be able to reach with the C$\mu$MD algorithm sufficiently low concentrations of NaOAc in solution. Each simulation box is comprised of a NaOAc polymorph I\cite{Hsu1983} crystal slab exposing face $\{200\}$ to the solution, whereas the top surface is comprised of an unfinished surface layer containing an unfinished row with kink sites. Figure \ref{fig:setups} shows visualizations of the simulation boxes of all six studied systems. The corresponding unfinished surface layers are shown in Figure \ref{fig:unfinishedl}. The unfinished surface layer was cut along the $[010]$ direction. Along this edge of the particular face $\{200\}$, Na$^+$ and AcO$^-$ are linearly arranged as dimers, which allows us to focus on the minimum of two types growth and dissolution processes. Namely one for each of the two ions, in order to extract the solute concentration dependent energy difference between grown and dissolved ion at the corresponding kink site, $\Delta F_{\text{Na}+}$ and $\Delta F_{\text{AcO}-}$. 
With their sum, $\Delta F = \Delta F_{\text{Na}+} + \Delta F_{\text{AcO}-}$, which corresponds to the energy difference for the dimeric unit, we can consequently identify the solubility as the mole fraction, $\chi$, where $\Delta F = 0$, as discussed in the main text.

\begin{table}[!htbp]
\caption{Simulation box specifications of the studied systems.}
\centering
\begin{tabular*}{\textwidth}{@{\extracolsep{\fill}}lccccccccc}
\toprule
                     & & pure MeOH  & & \multicolumn{3}{c}{MeOH-PrOH} & & \multicolumn{2}{c}{MeOH-MeCN} \\
\si{\percent}        & & 100        & & 80-20      & 60-40   & 40-60  & &  75-25        &  50-50        \\
\midrule
$N_\text{AcONa}$ [-] & & 139       & & 131     & 129     & 228       & & 129     & 228                 \\
$N_\text{MeOH}$ [-]  & & 580       & & 495     & 363     & 488       & & 446     & 646                 \\
$N_\text{MeCN}$ [-]  & &   0       & &   0     &   0     &   0       & & 116     & 504                 \\
$N_\text{PrOH}$ [-]  & &   0       & &  66     & 129     & 390       & &   0     &   0                 \\
$L_x$ [nm]           & & 2.48781   & & 2.48781 & 2.48781 & 3.73161   & & 2.48781 & 3.73161             \\
$L_y$ [nm]           & & 2.89733   & & 2.89733 & 2.89733 & 3.86232   & & 2.89733 & 3.86232             \\
$L_z$ [nm]           & & 7.12771   & & 7.42265 & 7.28108 & 7.14148   & & 7.32957 & 7.80798             \\
$T$ [K]              & & 300       & & 300     & 300     & 300       & & 300     & 300                 \\  
$p$ [bar]            & &   1       & &   1     &   1     &   1       & &   1     &   1                 \\
\bottomrule
\end{tabular*}
\label{tab:boxspecs}
\end{table}

We shall now explain how the reported simulation configurations were generated. Each equilibration simulation reported in this work was performed with Gromacs 2016.5 \cite{Abraham2015}. All simulations were run with full atomistic description, periodic boundary conditions, particle mesh Ewald approach \cite{Ewald1921,Darden1993} for the calculation of electrostatic interactions, where the non-bonded cutoff was set to 1 nm. The covalent bonds involving hydrogens were constrained at their equilibrium distances using the LINCS algorithm \cite{Hess2008a,Hess2008b}. To equilibrate the simulation boxes to the desired temperature and pressure, the velocity rescaling thermostat \cite{Bussi2009} and Parrinello-Rahman barostat \cite{Parrinello1981} were used respectively.

In a first step, the seed crystal of the NaOAc polymorph I was generated from XRD data \cite{Hsu1983} with face \{200\} perpendicular to the $z$ axis. The crystal system energy was then minimized with the conjugate gradient algorithm with a maximum force tolerance of 50 kJ mol$^{-1}$ nm$^{-1}$. The energy minimized crystal structure was then thermally equilibrated for 1 ns at $NVT$ conditions with a time integration step of 0.5 fs to reach the target temperature of $T$ = 300 K. The crystal system pressure was then equilibrated to the targeted 1 bar at $NPT$ conditions with the anisotropic Parrinello-Rahman barostat using a time integration step of 0.5 fs. After the pressure equilibration of 5 ns, the simulation was continued for further 20 ns to sample the average box lengths $L_x$ and $L_y$ and to identify the simulation frame closest to these average lengths, which was used as seed crystal in the following step.

The seed crystal with the averaged $L_x$ and $L_y$ box lengths was submerged in the particular NaOAc, solvent, and antisolvent solution using the Gromacs genbox utility \cite{Hess2008b}. We expose only face \{200\} to the solution. Face \{200\} is perpendicular to the $z$ axis of the simulation box. The simulation box energy minimization and temperature equilibration were performed in the same fashion as for the crystal system. The pressure equilibration was performed for 5 ns at $NPT$ conditions using the semi-isotropic Parrinello-Rahman barostat, where we allowed the box to change in size along the $z$ axis while keeping the already averaged $L_x$ and $L_y$ box lengths constant. The simulation was continued for another 20 ns, from which the average box length $L_z$ was sampled. The simulation frame closest to $L_z$ was used as initial condition for the final preparation step.

In the final step we have used the C$\mu$MD algorithm \cite{Perego2015} to generate the desired concentration profiles in the bulk liquid. Further details on the C$\mu$MD algorithm are discussed in the following Section \ref{sec:S3}.
During the solution concentration equilibration, a potential was introduced through adsorption site CVs (discussed in Section \ref{sec:S4.1}) which pushed all Na$^+$ and AcO$^-$ ions, which do not belong to the unfinished surface layer, away from the crystal surface. The unfinished surface layer ions were prevented from dissolving by the use of harmonic potentials introduced through surface structure CVs, which are discussed in Section \ref{sec:S4.2}. The center of mass of the two most inner layers was fixed with a harmonic potential to prevent the drift of the crystal. This drift would be undesirable for our setup. Simulation times of 50 ns ensure well equilibrated solute concentration profiles.

\section{Constant chemical potential method}\label{sec:S3}

To keep the chemical potential of NaOAc constant in the vicinity of the crystal surface comprising the kink site, the C$\mu$MD algorithm \cite{Perego2015} is introduced.

\begin{figure}[!htbp]
\begin{center}
\includegraphics[width=\textwidth]{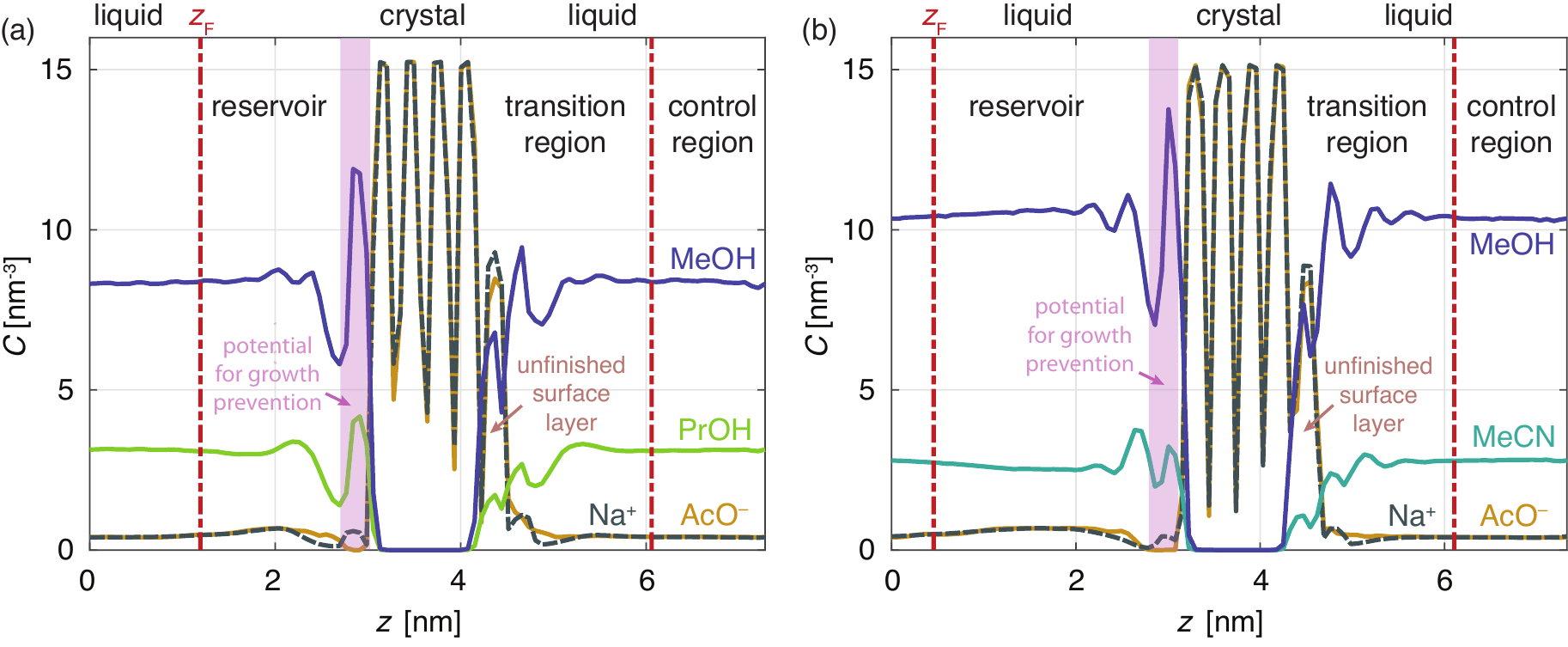}
\end{center}
\caption{Concentration profiles of each species, given in number of ions or molecules per nm$^{-3}$, along the $z$ axis for two representative simulation runs. a) NaOAc in 60-40\% MeOH-PrOH with an AcO$^-$ control region target concentration $C_0$ = 0.4 nm$^{-3}$. b) NaOAc in 75-25\% MeOH-MeCN with an AcO$^-$ control region target concentration $C_0$ = 0.5 nm$^{-3}$. The blue lines correspond to the MeOH, green lines to PrOH, turquoise lines to MeCN, brown lines to AcO$^-$ and dashed grey lines to Na$^+$ concentration profiles. The boundaries between the different liquid compartments, i.e. transition region, control region, and reservoir, are marked with red vertical dashed lines, whereat the position of the external force field is marked by $z_\text{F}$. The crystal with 4 fully grown layers (4 Na$^+$ and AcO$^-$ density peaks) is positioned in the center.
The unfinished surface layer is shown on the right side of the crystal with an AcO$^-$ peak of maximum concentration $C$ = 8 nm$^{-3}$. The action region of the potential, which prevents the opposite crystal surface from growing, is shaded in light purple. The concentration profiles were averaged over 1.2 $\mu$s of simulations.} \label{fig:CmuMD}
\end{figure}

We shall briefly discuss the function of the C$\mu$MD algorithm. As shown in Figure \ref{fig:CmuMD}, the liquid phase of the system is segmented into 3 different parts along the $z$ axis perpendicular to the crystal surface: the transition region, control region and reservoir. The transition region starts at the crystal-liquid interface and finishes where the concentration gradient of the solute becomes zero. The transition region is followed by the control region in which the C$\mu$MD algorithm tracks the concentration of solute molecules, $C_\text{CR}(t)$, at each time instant $t$. If $C_\text{CR}(t)$ is below the target concentration, $C_0$, an external force, $F^\mu_i$, acts at the interface between the control region and reservoir at position $z_\text{F}$. $F^\mu_i$ is defined as follows
\begin{equation}
F^\mu_i = k^\mu (C_\text{CR}(t) - C_0) G_\omega(z_i,z_\text{F}).
\end{equation}
$G_\omega(z_i,z_\text{F})$ is a bell shaped function
\begin{equation}
G_\omega(z_i,z_\text{F}) = \frac{1}{4\omega} \left[ 1 + \cosh \left( \frac{z_i - z_\text{F}}{\omega} \right) \right]^{-1}.
\end{equation}
$\omega$ defines the height and width of the bell curve.
$F^\mu_i$ will accelerate solute molecules from the reservoir towards the control region, if $C_\text{CR}$ is smaller than $C_0$. If $C_\text{CR}$ is larger than $C_0$, $F^\mu_i$ will accelerate the solute molecules out of the control region towards the reservoir. This ensures a constant solute concentration in the control region and allows us to run growth and dissolution simulations at constant chemical potential.

To prevent the growth of the crystal surface on the opposite site of the one containing the kink site, a harmonic potential is introduced (see Figure \ref{fig:CmuMD}). The potential acts through adsorption site CVs, which are discussed in Section \ref{sec:S4.1}, and prevents ions dissolved in the reservoir from adsorbing at the surface. At the same time, the surface layer is prevented from dissolving by the use of another harmonic potential and surface structure CVs, which are discussed in Section \ref{sec:S4.2}.
The crystal morphology is not affected by these potentials.

The C$\mu$MD algorithm was initially developed for a binary non-ionic molecular system and later extended to a three component aqueous rock salt solution\cite{Karmakar2019}.
In this work however we apply the algorithm to a quaternary liquid system involving two ions, a solvent and an antisolvent. Despite this more complicated liquid phase, the C$\mu$MD algorithm is able to keep the chemical potential constant as shown in Figures \ref{fig:CmuMD}a and \ref{fig:CmuMD}b showing the time averaged subspecies concentration profiles for the cases 60-40\% MeOH-PrOH and 75-25\% MeOH-MeCN respectively.
Even if the ions are mostly dissociated in the solution, we track and accelerate only the AcO$^-$ with the C$\mu$MD algorithm, since the Na$^+$ are following the diffusion movement of AcO$^-$ due to the coulombic attraction of the ions. In both Figures \ref{fig:CmuMD}a and \ref{fig:CmuMD}b the concentration curves of AcO$^-$ and Na$^+$ overlap.
The targeted concentration profiles in the control region are met for both the NaOAc concentration, as well as for the desired solvent-antisolvent concentration ratio with excellent accuracy.
The C$\mu$MD parameters used for each simulation setup are reported in Table \ref{tab:CmuMD}.

\begin{table}[!htbp]
\caption{Values of the C$\mu$MD parameters. Length parameters are given in fractional coordinates.}
\centering
\begin{tabular*}{\textwidth}{@{\extracolsep{\fill}}lccccccccc}
\toprule
                      & & pure MeOH & & \multicolumn{3}{c}{MeOH-PrOH} & & \multicolumn{2}{c}{MeOH-MeCN} \\
\si{\percent}         & & 100       & & 80-20   & 60-40   & 40-60     & &  75-25  &  50-50              \\
\midrule
$\omega/L_z$ [-]      & & 0.02      & & 0.02    & 0.02    & 0.02      & & 0.02    & 0.02                \\
$z_\text{TR}/L_z$ [-] & & 0.20      & & 0.24    & 0.24    & 0.24      & & 0.20    & 0.20                \\
$z_\text{CR}/L_z$ [-] & & 0.30      & & 0.33    & 0.33    & 0.31      & & 0.30    & 0.33                \\
$z_\text{F}/L_z$ [-]  & & 0.52      & & 0.59    & 0.59    & 0.57      & & 0.52    & 0.55                \\
$\Delta z/L_z$ [-]    & & 1/120     & & 1/120   & 1/120   & 1/120     & & 1/120   & 1/120               \\
\bottomrule
\end{tabular*}
\label{tab:CmuMD}
\end{table}

\section{Collective variables (CVs)}\label{sec:S4}

\subsection{Adsorption site CVs}\label{sec:S4.1} 

In the biased simulation runs we aim to sample many growth and dissolution events of the biased kink site to obtain the needed convergence accuracy. As discussed in Section \ref{sec:S5}, the time to acquire enough growth and dissolution events is in the order of 1.5 $\mu$s for the studied NaOAc systems.
Although the growth of NaOAc at kinks and edges of the unfinished surface layer is slow in unbiased simulations, in a time span of 1.5 $\mu$s this process is highly likely occur. And this growth at other adsorption sites around the biased kink site can disrupt and slow down the sampling performance of the biased simulation.

To prevent the growth of NaOAc at other sites apart from the biased kink site, we introduce harmonic potential walls through adsorption site CVs. We use two types of adsorption site CVs, namely spherical ones for single adsorption sites such as kinks and prismatic ones for edges. The spherical adsorption site CVs are defined with logistic step functions for a particular adsorption site $a$ and ion type as follows
\begin{equation}
s_{a,\text{ion}} = \sum_i \left(1 - \frac{1}{1+\exp(-\sigma_a(|\mathbf{r}_{i}-\mathbf{r}_a|-d_a))}\right),
\end{equation}
where $\sigma_a$ and $d_a$ define the steepness of the step and radius of the adsorption site respectively, $\mathbf{r}_{i}$ is the position of ion $i$, and $\mathbf{r}_a$ corresponds to the position of adsorption site $a$.

The edge adsorption site CVs are also comprised of logistic switching functions and have following functional form for a particular ion type
\begin{align}
\begin{split}
s_{\varsigma,\text{ion}} = \sum_i \bigg[ \frac{1}{1 + \exp (-\sigma_\varsigma (x_i - x_\text{l}))} &\left( 1 - \frac{1}{1 + \exp (-\sigma_\varsigma (x_i - x_\text{u}))} \right) \\
\cdot \frac{1}{1 + \exp (-\sigma_\varsigma (z_i - z_\text{l}))} &\left( 1 - \frac{1}{1 + \exp (-\sigma_\varsigma (z_i - z_\text{u}))} \right) \bigg].
\end{split}
\end{align}
$\sigma_\varsigma$ dictates the steepness of the step functions. $x_i$ and $z_i$ are the positions of ion $i$ along the $x$ and $z$ axis. $x_l$ and $x_u$, as well as $z_l$ and $z_u$ are intervals which confine the adsorption site CV's region of action along the $x$ and $z$ axis.

Harmonic wall potentials are introduced through the spherical adsorption site CVs
\begin{equation}
V_{a,\text{ion}} =
\begin{cases}
0, & \text{if}\ s_a < s_{a,0}, \\
k_a (s_a - s_{a,0})^2, & \text{else},
\end{cases}
\end{equation}
and edge adsorption site CVs
\begin{equation}
V_{\varsigma,\text{ion}} =
\begin{cases}
0, & \text{if}\ s_\varsigma < s_{\varsigma,0}, \\
k_\varsigma (s_\varsigma - s_{\varsigma,0})^2, & \text{else},
\end{cases}
\end{equation}
where $k_a$ and $k_\varsigma$ are the force constants and $s_{a,0}$ and $s_{\varsigma,0}$ are the threshold values of the adsorption site CVs beyond which the bias potentials start to act.

\begin{figure}[!htbp]
\begin{center}
\includegraphics[width=\textwidth]{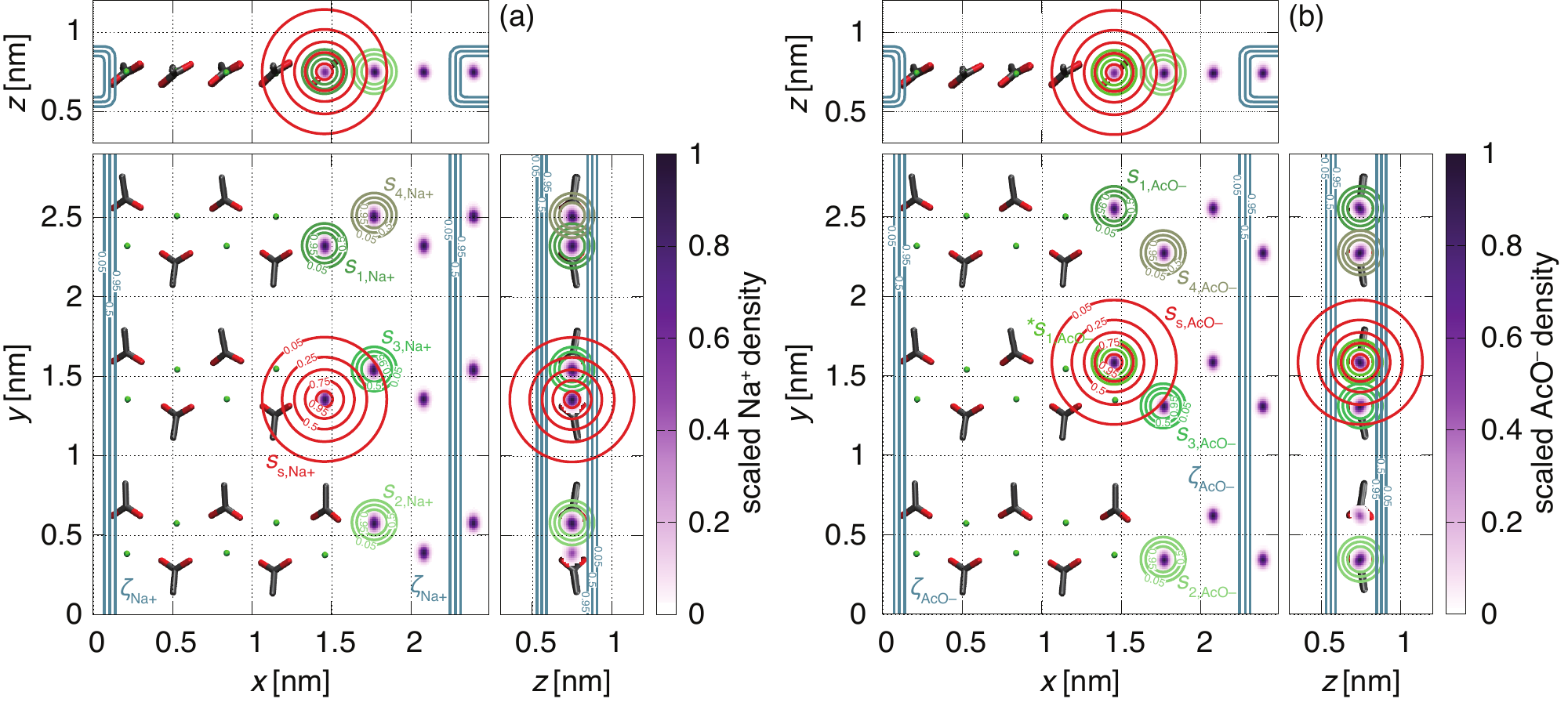}
\end{center}
\caption{Contour lines of the adsorption site CVs (green lines) shown for a) adsorption sites of Na$^+$ and b) adsorption sites of AcO$^-$ on the crystal surface comprising the biased kink site. The adsorption site CVs are the ones used for the pure MeOH solution simulation setup and their contour lines are shown at values of 0.01, 0.5, and 0.99.
Contour lines of the Gaussian like functions, $s_{\text{s},\text{Na+}}$ in a) and $s_{\text{s},\text{s},\text{AcO-}}$ in b), which are the kink site ion density terms in the biased CVs $s_{\text{b},\text{Na+}}$ and $s_{\text{b},\text{AcO-}}$ respectively, are colored in red.
The contours are shown as projections along all three coordinates. The unfinished surface layer ions are presented with structural formulas using the same color code as in Figure \ref{fig:setups} (hydrogens are omitted for clarity). The histograms belong to the scaled crystalline ion densities, a) Na$^+$ and b) AcO$^-$, of the surface layer which are not part of the unfinished surface layer and are sampled from 50 ns of unbiased simulations. These NaOAc crystalline positions overlap with the adsorption sites. The AcO$^-$ ion positions are defined as the center of mass of their two oxygens. The histograms were scaled by their maximum density values.
} \label{fig:adsorptionsites}
\end{figure}

The adsorption site CV parameters used for each simulation setup are listed in Table \ref{tab:adsNa} for the Na$^+$ ions and Table \ref{tab:adsAcO} for the AcO$^-$ ions. Spherical adsorption sites are enumerated from $a=1$ to 4.
The center of mass of the oxygen atoms O1 and O2 are defined as the AcO$^-$ ion positions.
The adsorption site CVs with the reported parameters are positioned along all edges and kinks except for the biased kink site.
These configurations of potentials allow us to keep the biased kink site environment intact for the time span necessary to obtain converged free energy profiles. The force constants of the potentials are set as mild as possible in order not to disturb ions from diffusing through these sites.
Figure \ref{fig:adsorptionsites} shows the contour lines of the adsorption site CVs, $s_{1,\text{Na+}}$, $s_{2,\text{Na}+}$, $s_{3,\text{Na}+}$, $s_{4,\text{Na}+}$ $s_{\varsigma,\text{Na}+}$, $s_{1,\text{AcO}-}$, $s_{2,\text{AcO}-}$, $s_{3,\text{AcO}-}$, $s_{4,\text{AcO}-}$, and $s_{\varsigma,\text{AcO}-}$, 
used for the pure MeOH solution simulation setup.

\begin{table}
\small
\caption{Values of adsorption site CVs for Na$^+$. The coordinate origin of the $z$ axis is set to the box center.}  \label{tab:adsNa}
\centering
\begin{tabular*}{\textwidth}{@{\extracolsep{\fill}}llccccccccc}
\toprule
&                   & & pure MeOH  & & \multicolumn{3}{c}{MeOH-PrOH} & & \multicolumn{2}{c}{MeOH-MeCN} \\
&\si{\percent}      & & 100        & & 80-20      & 60-40   & 40-60  & &  75-25        &  50-50        \\
\midrule
\multirow{7}{*}{$s_{1,\text{Na}^+}$}
& $r_a^{(x)}$ [nm]  & & 1.4584  & & 1.4576  & 1.4567  & 1.6569  & & 1.4587  & 1.3433  \\
& $r_a^{(y)}$ [nm]  & & 2.3234  & & 2.3232  & 0.5792  & 2.3287  & & 2.3242  & 1.0640  \\
& $r_a^{(z)}$ [nm]  & & 0.7506  & & 0.6860  & 0.7562  & 0.7514  & & 0.8713  & 0.7505  \\
& $\sigma_a$ [-]    & & 100     & & 100     & 100     & 100     & & 100     & 100     \\
& $d_a$ [nm]        & & 0.11    & & 0.11    & 0.11    & 0.11    & & 0.11    & 0.11    \\
& $k_a$ [kJ/mol]    & & 40      & & 40      & 40      & 40      & & 40      & 40      \\
& $s_{a,0}$ [-]     & & 0.1     & & 0.1     & 0.1     & 0.1     & & 0.1     & 0.1     \\
\midrule
\multirow{7}{*}{$s_{2,\text{Na}^+}$}
& $r_a^{(x)}$ [nm]  & & 1.7687  & & 1.7702  & 1.7676  & 1.6569  & & 1.7690  & 1.3433  \\
& $r_a^{(y)}$ [nm]  & & 0.5786  & & 0.5792  & 0.3908  & 3.2839  & & 0.5785  & 0.0915  \\
& $r_a^{(z)}$ [nm]  & & 0.7506  & & 0.6860  & 0.7562  & 0.7514  & & 0.8713  & 0.7505  \\
& $\sigma_a$ [-]    & & 100     & & 100     & 100     & 100     & & 100     & 100     \\
& $d_a$ [nm]        & & 0.11    & & 0.11    & 0.11    & 0.11    & & 0.11    & 0.11    \\
& $k_a$ [kJ/mol]    & & 20      & & 20      & 20      & 20      & & 20      & 20      \\
& $s_{a,0}$ [-]     & & 0.1     & & 0.1     & 0.1     & 0.1     & & 0.1     & 0.1     \\
\midrule
\multirow{7}{*}{$s_{3,\text{Na}^+}$}
& $r_a^{(x)}$ [nm]  & & 1.7687  & & 1.7702  & 1.7676  & 1.3455  & & 1.7690  & 1.0329  \\
& $r_a^{(y)}$ [nm]  & & 1.5442  & & 1.5439  & 1.3590  & 0.5666  & & 1.5430  & 2.8151  \\
& $r_a^{(z)}$ [nm]  & & 0.7506  & & 0.6860  & 0.7562  & 0.7514  & & 0.8713  & 0.7505  \\
& $\sigma_a$ [-]    & & 100     & & 100     & 100     & 100     & & 100     & 100     \\
& $d_a$ [nm]        & & 0.11    & & 0.11    & 0.11    & 0.11    & & 0.11    & 0.11    \\
& $k_a$ [kJ/mol]    & & 20      & & 20      & 20      & 20      & & 20      & 20      \\
& $s_{a,0}$ [-]     & & 0.1     & & 0.1     & 0.1     & 0.1     & & 0.1     & 0.1     \\
\midrule
\multirow{7}{*}{$s_{4,\text{Na}^+}$}
& $r_a^{(x)}$ [nm]  & & 1.7687  & & 1.7702  & 1.7676  & 1.3455  & & 1.7690  & -       \\
& $r_a^{(y)}$ [nm]  & & 2.5123  & & 2.5090  & 2.3227  & 3.4620  & & 2.5098  & -       \\
& $r_a^{(z)}$ [nm]  & & 0.7506  & & 0.6860  & 0.7562  & 0.7514  & & 0.8713  & -       \\
& $\sigma_a$ [-]    & & 100     & & 100     & 100     & 100     & & 100     & -       \\
& $d_a$ [nm]        & & 0.11    & & 0.11    & 0.11    & 0.11    & & 0.11    & -       \\
& $k_a$ [kJ/mol]    & & 20      & & 20      & 20      & 20      & & 20      & -       \\
& $s_{a,0}$ [-]     & & 0.1     & & 0.1     & 0.1     & 0.1     & & 0.1     & -       \\
\midrule
\multirow{7}{*}{$\varsigma_{\text{Na}^+}$}
& $\sigma_\varsigma$ [-]  & & 100     & & 100     & 100     & 100     & & 100     & 100     \\
& $x_\text{l}$ [nm]       & & 1.0371  & & 1.0361  & 1.0365  & 1.1753  & & 1.0369  & 0.8539  \\
& $x_\text{u}$ [nm]       & & 1.3471  & & 1.3461  & 1.3465  & 1.5053  & & 1.3469  & 1.1739  \\
& $z_\text{l}$ [nm]       & & 0.5600  & & 0.4960  & 0.5662  & 0.5584  & & 0.6813  & 0.5606  \\
& $z_\text{u}$ [nm]       & & 0.8800  & & 0.8060  & 0.8762  & 0.8884  & & 0.9913  & 0.8906  \\
& $k_\varsigma$ [kJ/mol]  & & 25      & & 25      & 25      & 25      & & 25      & 25      \\
& $\varsigma_0$ [nm]      & & 0.5     & & 0.5     & 0.5     & 0.5     & & 0.5     & 0.5     \\
\bottomrule
\end{tabular*}
\end{table}

\begin{table}
\small
\caption{Values of adsorption site CVs for AcO$^-$. The coordinate origin of the $z$ axis is set to the box center.} \label{tab:adsAcO}
\centering
\begin{tabular*}{\textwidth}{@{\extracolsep{\fill}}llccccccccc}
\toprule
&                   & & pure MeOH  & & \multicolumn{3}{c}{MeOH-PrOH} & & \multicolumn{2}{c}{MeOH-MeCN} \\
&\si{\percent}      & & 100        & & 80-20      & 60-40   & 40-60  & &  75-25        &  50-50        \\
\midrule
\multirow{7}{*}{$s_{0,\text{AcO}^-}$}
& $r_a^{(x)}$ [nm]       & & 1.4571  & & 1.4576  & 1.4559  & 1.6577  & & 1.4583  & 1.3449  \\
& $r_a^{(y)}$ [nm]       & & 1.5908  & & 1.5911  & 1.3103  & 1.5981  & & 1.5916  & 1.7992  \\
& $r_a^{(z)}$ [nm]       & & 0.7442  & & 0.6794  & 0.7496  & 0.7484  & & 0.8639  & 0.7476  \\
& $\sigma_a$ [-]         & & 100     & & 100     & 100     & 100     & & 100     & 100     \\
& $d_a$ [nm]             & & 0.11    & & 0.11    & 0.11    & 0.11    & & 0.11    & 0.11    \\
& $k_a$ [kJ/mol]         & & 40      & & 40      & 40      & 40      & & 40      & 40      \\
& $s_{a,0}$ [-]          & & 0.1     & & 0.1     & 0.1     & 0.1     & & 0.1     & 0.1     \\
\midrule
\multirow{7}{*}{$s_{1,\text{AcO}^-}$}
& $r_a^{(x)}$ [nm]       & & 1.4571  & & 1.4576  & 1.4559  & 1.6577  & & 1.4583  & 1.3449  \\
& $r_a^{(y)}$ [nm]       & & 2.5567  & & 2.5567  & 0.3459  & 2.5621  & & 2.5573  & 3.7210  \\
& $r_a^{(z)}$ [nm]       & & 0.7442  & & 0.6794  & 0.7496  & 0.7484  & & 0.8639  & 0.7476  \\
& $\sigma_a$ [-]         & & 100     & & 100     & 100     & 100     & & 100     & 100     \\
& $d_a$ [nm]             & & 0.11    & & 0.11    & 0.11    & 0.11    & & 0.11    & 0.11    \\
& $k_a$ [kJ/mol]         & & 40      & & 40      & 40      & 40      & & 40      & 40      \\
& $s_{a,0}$ [-]          & & 0.1     & & 0.1     & 0.1     & 0.1     & & 0.1     & 0.1     \\
\midrule
\multirow{7}{*}{$s_{2,\text{AcO}^-}$}
& $r_a^{(x)}$ [nm]       & & 1.7674  & & 1.7683  & 1.7682  & 1.6577  & & 1.7687  & 1.3449  \\
& $r_a^{(y)}$ [nm]       & & 0.3452  & & 0.3459  & 0.6245  & 3.5243  & & 0.3452  & 0.8306  \\
& $r_a^{(z)}$ [nm]       & & 0.7442  & & 0.6794  & 0.7496  & 0.7484  & & 0.8639  & 0.7476  \\
& $\sigma_a$ [-]         & & 100     & & 100     & 100     & 100     & & 100     & 100     \\
& $d_a$ [nm]             & & 0.11    & & 0.11    & 0.11    & 0.11    & & 0.11    & 0.11    \\
& $k_a$ [kJ/mol]         & & 20      & & 20      & 20      & 20      & & 20      & 20      \\
& $s_{a,0}$ [-]          & & 0.1     & & 0.1     & 0.1     & 0.1     & & 0.1     & 0.1     \\
\midrule
\multirow{7}{*}{$s_{3,\text{AcO}^-}$}
& $r_a^{(x)}$ [nm]       & & 1.7674  & & 1.7683  & 1.7682  & 1.3462  & & 1.7687  & 1.0343  \\
& $r_a^{(y)}$ [nm]       & & 1.3125  & & 1.3095  & 1.5934  & 1.3102  & & 1.3097  & 2.0772  \\
& $r_a^{(z)}$ [nm]       & & 0.7442  & & 0.6794  & 0.7496  & 0.7484  & & 0.8639  & 0.7476  \\
& $\sigma_a$ [-]         & & 100     & & 100     & 100     & 100     & & 100     & 100     \\
& $d_a$ [nm]             & & 0.11    & & 0.11    & 0.11    & 0.11    & & 0.11    & 0.11    \\
& $k_a$ [kJ/mol]         & & 20      & & 20      & 20      & 20      & & 20      & 20      \\
& $s_{a,0}$ [-]          & & 0.1     & & 0.1     & 0.1     & 0.1     & & 0.1     & 0.1     \\
\midrule
\multirow{7}{*}{$s_{4,\text{AcO}^-}$}
& $r_a^{(x)}$ [nm]       & & 1.7674  & & 1.7683  & 1.7682  & 1.3462  & & 1.7687  & -       \\
& $r_a^{(y)}$ [nm]       & & 2.2782  & & 2.2764  & 2.5563  & 0.3431  & & 2.2767  & -       \\
& $r_a^{(z)}$ [nm]       & & 0.7442  & & 0.6794  & 0.7496  & 0.7484  & & 0.8639  & -       \\
& $\sigma_a$ [-]         & & 100     & & 100     & 100     & 100     & & 100     & -       \\
& $d_a$ [nm]             & & 0.11    & & 0.11    & 0.11    & 0.11    & & 0.11    & -       \\
& $k_a$ [kJ/mol]         & & 20      & & 20      & 20      & 20      & & 20      & -       \\
& $s_{a,0}$ [-]          & & 0.1     & & 0.1     & 0.1     & 0.1     & & 0.1     & -       \\
\midrule
\multirow{7}{*}{$\varsigma_{\text{AcO}^-}$}
& $\sigma_\varsigma$ [-] & & 100     & & 100     & 100     & 100     & & 100     & 100     \\
& $x_\text{l}$ [nm]      & & 1.0357  & & 1.0359  & 1.0361  & 1.1753  & & 1.0362  & 0.8339  \\
& $x_\text{u}$ [nm]      & & 1.3457  & & 1.3459  & 1.3461  & 1.5053  & & 1.3462  & 1.1540  \\
& $z_\text{l}$ [nm]      & & 0.5542  & & 0.4894  & 0.5596  & 0.5584  & & 0.6739  & 0.5542  \\
& $z_\text{u}$ [nm]      & & 0.8642  & & 0.7994  & 0.8696  & 0.8884  & & 0.9839  & 0.8642  \\
& $k_\varsigma$ [kJ/mol] & & 25      & & 25      & 25      & 25      & & 25      & 25      \\
& $\varsigma_0$ [nm]     & & 0.5     & & 0.5     & 0.5     & 0.5     & & 0.5     & 0.5     \\
\bottomrule
\end{tabular*}
\end{table}

\subsection{Surface structure CVs}\label{sec:S4.2}  

During the NaOAc kink growth and dissolution simulations, it is likely that ions of crystal layers, which are exposed to solution, dissolve and alter the environment of the biased kink site. That would disrupt the sampling of the growth and dissolution events at the particular kink site.

To prevent such undesired dissolution of ions, that belong to crystal layers at the crystal-solution interface, harmonic potential walls are introduced for each of these layers, $l$, and for each ion type through surface structure CVs, $s_{\text{st},l,\text{ion}}$. $s_{\text{st},l,\text{ion}}$ is defined as a logistic function
\begin{equation}
s_{\text{st},l,\text{ion}} = \frac{1}{1 + \exp{(-\sigma_\text{st}(\tilde{s}_\text{st}-\tilde{s}_\text{st,0}))}}, 
\end{equation}\label{eq:s_st}
where $\tilde{s}_{\text{st},0}$ denotes the step position and $\sigma_\text{st}$ the steepness of following function
\begin{equation}
\tilde{s}_{\text{st},\text{ion}} = \sum_i \left( \sum_k \left[\cos^{\eta_x} \left( \frac{\nu_x \pi}{L_x} (x_i - \bar{x}_k) \right) \cos^{\eta_y} \left( \frac{\nu_y \pi}{L_y} (y_i - \bar{y}_k) \right)\right] \exp \left\{ -\frac{(z_i-\bar{z}_k)^2}{2 \sigma_z^2} \right\} \right).
\end{equation}
Along the $xy$ plane we introduce sinusoidal functions which capture the lattice positions of the ions. $\eta_x$ and $\eta_y$ define the widths of the peaks and need to be even positive integers. $\nu_x$ and $\nu_y$ correspond to the number of unit cells along the $x$ and $y$ axis. $L_x$ and $L_y$ are the simulation box lengths in the $x$ and $y$ dimensions.
$\bar{x}_k$ and $\bar{y}_k$ is the $k$-th ion position in the $xy$ unit cell plane.
Along the $z$ axis of the simulation box, where we have no continuous crystal periodicity, we introduce Gaussian functions. $\bar{z}_k$ defines the average position of ion $k$ in $z$ direction of the simulation box and $\sigma_z$ defines the width of the Gaussian functions. The terms are summed over each ion $i$ that belongs to the corresponding layer $l$.

Harmonic wall potentials are introduced for each $s_{\text{st},l,\text{ion}}$ in the following form
\begin{equation}
V_{\text{st},l,\text{ion}} =
\begin{cases}
k_\text{st} (s_\text{st} - s_\text{st,0})^2, & \text{if}\ s_\text{st} < s_\text{st,0}, \\
0, & \text{else}.
\end{cases}
\end{equation}
$k_\text{st}$ is the force constant and $s_{\text{st},0}$ denotes the threshold value below which the harmonic potential starts to act.

\begin{figure}[!htbp]
\begin{center}
\includegraphics[width=\textwidth]{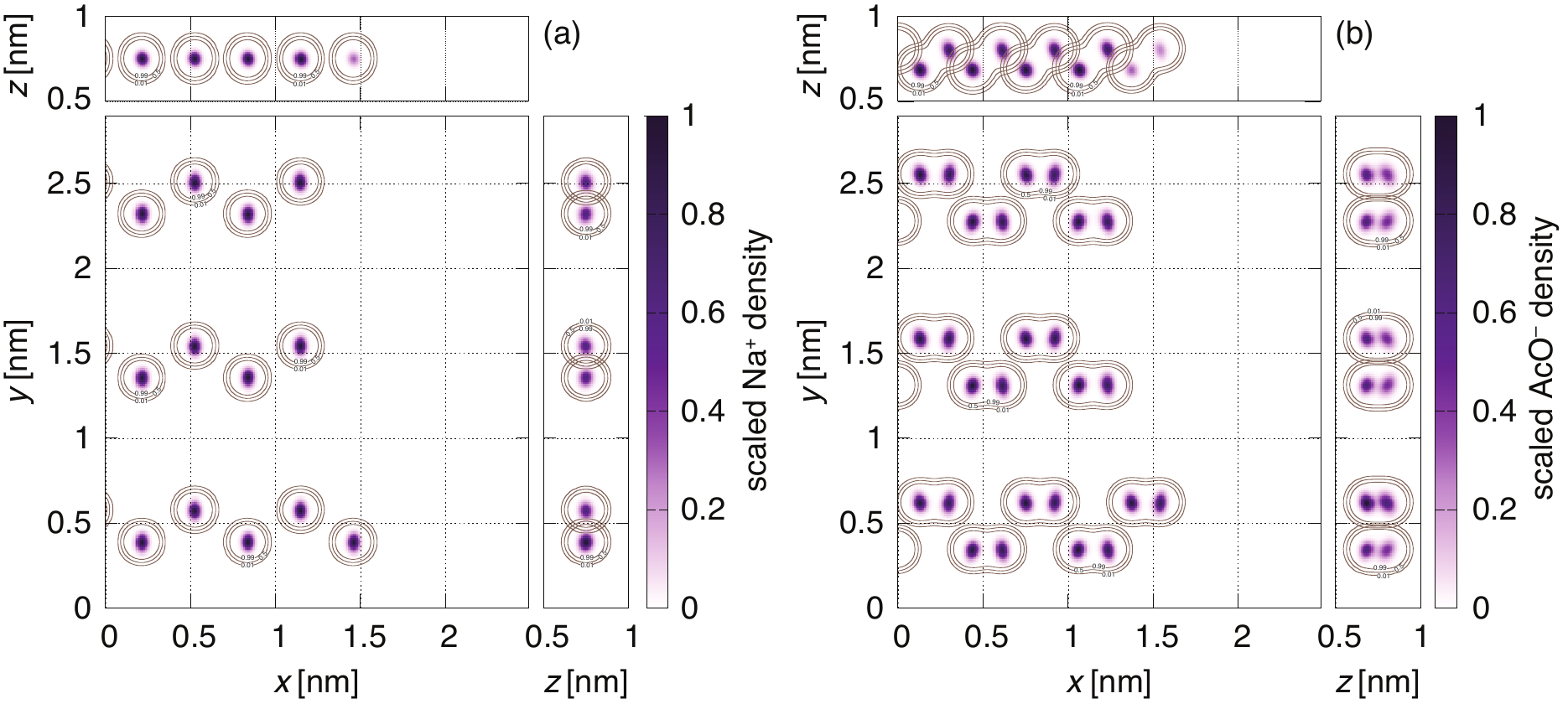}
\end{center}
\caption{Contour lines of the surface structure CVs (brown lines) for a) Na$^+$ and b) AcO$^-$ used in the pure MeOH simulation setups. 
The contour lines are shown at values of 0.01 (most outer rings), 0.5, and 0.99 (most inner rings).
The histograms correspond to the scaled densities of a) Na$^+$ ions and b) AcO$^-$ oxygens which belong to the unfinished surface layer. The histograms were sampled from 50 ns of unbiased simulation and scaled by their maximum density values.} \label{fig:surfacestructure}
\end{figure}

The shape of $s_{\text{st},l,\text{ion}}$ is chosen such that ions in the given crystalline layer are not affected by the harmonic potential as long as the ions are within their crystal lattice position. This is the case where $s_{\text{st},l,\text{ion}}$ = 1 for the given ion $i$. $s_{\text{st},l,\text{ion}}$ should have values below 1 only when the ion $i$ is outside its lattice position. It is important that $V_{\text{st},l,\text{ion}}$ does not interfere with the natural lattice vibrations of the crystal. This is achieved by setting the parameters of $s_{\text{st},l,\text{ion}}$ such that for each ion $i$ at its vibration amplitude position $s_{\text{st},l,\text{ion}}$ still has a value of 1.
If the vibrations are hindered, the simulations give distorted growth and dissolution processes.
The vibration amplitudes can be obtained from unbiased simulations of the crystal exposed to solution. Figures \ref{fig:surfacestructure}a and \ref{fig:surfacestructure}b show the contour lines of $s_{\text{st},l,\text{ion}}$ together with the histograms of the ion positions of the unfinished surface layer for Na$^+$ and AcO$^-$. For AcO$^-$, the oxygen positions are considered. The contour lines show, that $s_{\text{st},l,\text{ion}}$ only has values below 1 at a distance where the ion density approaches values of zero.

In the biased simulations we applied potential walls to the unfinished surface layer, $s_{\text{st},5,\text{ion}}$, the layer adjacent to the unfinished surface layer, $s_{\text{st},4,\text{ion}}$, and the layer on the opposite crystal surface, $s_{\text{st},1,\text{ion}}$. The appropriate parameter values are listed in Tables \ref{tab:sstNa5}, \ref{tab:sstAcO5}, \ref{tab:sstNa4}, \ref{tab:sstAcO4}, \ref{tab:sstNa1}, and \ref{tab:sstAcO1}.

\begin{table}[!htbp]
\caption{Values of the surface structure CV parameters for Na$^+$ of the unfinished surface layer. $^{*)}$In the Na$^+$ kink growth simulations, a $s_{\text{st},0}$ value of 13 or 17 was used. For kink growth simulations of AcO$^-$, the Na$^+$ of the biased dimeric unit was kept in its crystalline position with the surface structure CV using a $s_{\text{st},0}$ value of 14 or 18.}
\centering
\begin{tabular*}{\textwidth}{@{\extracolsep{\fill}}lccccccccc}
\toprule
                   & & pure MeOH  & & \multicolumn{3}{c}{MeOH-PrOH} & & \multicolumn{2}{c}{MeOH-MeCN} \\
\si{\percent}      & & 100        & & 80-20      & 60-40   & 40-60  & &  75-25        &  50-50        \\
\midrule
$\nu_x$ [-]            & & 4       & & 4       & 4       & 6       & & 4       & 6       \\
$L_x$ [nm]             & & 2.48781 & & 2.48781 & 2.48781 & 3.73161 & & 2.48781 & 3.73161 \\
$\eta_x$ [-]           & & 10      & & 10      & 10      & 10      & & 10      & 10      \\
$\bar{x}_1$ [nm]       & & -0.7199 & & -0.7184 & -0.7184 & -0.2089 & & -0.7188 & 2.2765  \\
$\bar{x}_2$ [nm]       & & -0.4080 & & -0.4080 & -0.4080 & 0.1017  & & -0.4069 & 2.5884  \\
$\nu_y$ [-]            & & 3       & & 3       & 3       & 4       & & 3       & 4       \\
$L_y$ [nm]             & & 2.89733 & & 2.89733 & 2.89733 & 3.86232 & & 2.89733 & 3.86232 \\
$\eta_y$ [-]           & & 26      & & 26      & 26      & 26      & & 26      & 26      \\
$\bar{y}_1$ [nm]       & & -1.0579 & & -0.8695 & -0.8695 & -3.4640 & & -0.8699 & -2.9848 \\
$\bar{y}_2$ [nm]       & & -0.8694 & & -1.0572 & -1.0572 & -3.2855 & & -1.0580 & -2.7983 \\
$\bar{z}_1$ [nm]       & & 0.7562  & & 0.6860  & 0.6860  & 0.7514  & & 0.8713  & 0.7506  \\
$\bar{z}_2$ [nm]       & & 0.7562  & & 0.6860  & 0.6860  & 0.7514  & & 0.8713  & 0.7506  \\
$\sigma_z$ [nm]        & & 0.06    & & 0.06    & 0.06    & 0.06    & & 0.06    & 0.06    \\
$\sigma_\text{st}$ [-] & & 150     & & 150     & 150     & 150     & & 150     & 150     \\
$\tilde{s}_{\text{st},0}$ [-] & & 0.10      & & 0.10      & 0.10      & 0.10      & & 0.10      & 0.10 \\
$k_\text{st}$ [kJ/mol]        & & 25        & & 25        & 25        & 25        & & 25        & 25 \\
$s_{\text{st},0}$ [-]         & & 13/14$^*$ & & 13/14$^*$ & 13/14$^*$ & 17/18$^*$ & & 13/14$^*$ & 17/18$^*$ \\
\bottomrule
\end{tabular*}
\label{tab:sstNa5}
\end{table}

\begin{table}[!htbp]
\caption{Values of the surface structure CV parameters for AcO$^-$ of the unfinished surface layer.}
\centering
\begin{tabular*}{\textwidth}{@{\extracolsep{\fill}}lccccccccc}
\toprule
                   & & pure MeOH  & & \multicolumn{3}{c}{MeOH-PrOH} & & \multicolumn{2}{c}{MeOH-MeCN} \\
\si{\percent}      & & 100        & & 80-20      & 60-40   & 40-60  & &  75-25        &  50-50        \\
\midrule
$\nu_x$ [-]            & & 4       & & 4       & 4       & 6       & & 4       & 6       \\
$L_x$ [nm]             & & 2.48781 & & 2.48781 & 2.48781 & 3.73161 & & 2.48781 & 3.73161 \\
$\eta_x$ [-]           & & 12      & & 12      & 12      & 12      & & 12      & 12      \\
$\bar{x}_1$ [nm]       & & -0.9452 & & -0.9439 & -0.9439 & -0.2943 & & -0.9446 & -1.5411 \\
$\bar{x}_2$ [nm]       & & -0.8049 & & -0.8037 & -0.8037 & -0.1228 & & -0.8018 & -1.3676 \\
$\bar{x}_3$ [nm]       & & -0.6332 & & -0.6339 & -0.6339 & 0.0161  & & -0.6323 & -1.2288 \\
$\bar{x}_4$ [nm]       & & -0.4927 & & -0.4926 & -0.4926 & 0.1887  & & -0.4931 & -1.0572 \\
$\nu_y$ [-]            & & 3       & & 3       & 3       & 4       & & 3       & 4       \\
$L_y$ [nm]             & & 2.89733 & & 2.89733 & 2.89733 & 3.86232 & & 2.89733 & 3.86232 \\
$\eta_y$ [-]           & & 30      & & 30      & 30      & 30      & & 30      & 30      \\
$\bar{y}_1$ [nm]       & & -0.8256 & & -0.8244 & -0.8244 & -0.3335 & & -0.8244 & 0.1454  \\
$\bar{y}_2$ [nm]       & & -0.1365 & & -1.1030 & -1.1030 & -0.3335 & & -1.1039 & 0.1454  \\
$\bar{y}_3$ [nm]       & & -0.1365 & & -1.1030 & -1.1030 & -0.6212 & & -1.1039 & -0.1288 \\
$\bar{y}_4$ [nm]       & & -0.8256 & & -0.8244 & -0.8244 & -0.6212 & & -0.8244 & -0.1288 \\
$\bar{z}_1$ [nm]       & & 0.8109  & & 0.7409  & 0.7409  & 0.8090  & & 0.9290  & 0.8092  \\
$\bar{z}_2$ [nm]       & & 0.6929  & & 0.6238  & 0.6238  & 0.6960  & & 0.8122  & 0.6915  \\
$\bar{z}_3$ [nm]       & & 0.8109  & & 0.7409  & 0.7409  & 0.8090  & & 0.9290  & 0.8092  \\
$\bar{z}_4$ [nm]       & & 0.6929  & & 0.6238  & 0.6238  & 0.6960  & & 0.8122  & 0.6915  \\
$\sigma_z$ [nm]        & & 0.06    & & 0.06    & 0.06    & 0.06    & & 0.06    & 0.06    \\
$\sigma_\text{st}$ [-] & & 0.10    & & 0.10    & 0.10    & 0.10    & & 0.10    & 0.10    \\
$\tilde{s}_{\text{st},0}$ [-] & & 150 & & 150 & 150 & 150 & & 150 & 150 \\
$k_\text{st}$ [kJ/mol]        & & 25  & & 25  & 25  & 25  & & 25  & 25  \\
$s_{\text{st},0}$ [-]         & & 26  & & 26  & 26  & 34  & & 26  & 34  \\
\bottomrule
\end{tabular*}
\label{tab:sstAcO5}
\end{table}

\begin{table}[!htbp]
\caption{Values of the surface structure CV parameters for Na$^+$ of layer 4.}
\centering
\begin{tabular*}{\textwidth}{@{\extracolsep{\fill}}lccccccccc}
\toprule
                   & & pure MeOH  & & \multicolumn{3}{c}{MeOH-PrOH} & & \multicolumn{2}{c}{MeOH-MeCN} \\
\si{\percent}      & & 100        & & 80-20      & 60-40   & 40-60  & &  75-25        &  50-50        \\
\midrule
$\nu_x$ [-]            & & 4       & & 4       & 4       & 6       & & 4       & 6       \\
$L_x$ [nm]             & & 2.48781 & & 2.48781 & 2.48781 & 3.73161 & & 2.48781 & 3.73161 \\
$\eta_x$ [-]           & & 10      & & 10      & 10      & 10      & & 10      & 10      \\
$\bar{x}_1$ [nm]       & & -0.8335 & & -0.8341 & -0.8341 & -1.6510 & & -0.8328 & -1.6517 \\
$\bar{x}_2$ [nm]       & & -0.5232 & & -0.5225 & -0.5225 & -1.3424 & & -0.5219 & -1.3398 \\
$\nu_y$ [-]            & & 3       & & 3       & 3       & 4       & & 3       & 4       \\
$L_y$ [nm]             & & 2.89733 & & 2.89733 & 2.89733 & 3.86232 & & 2.89733 & 3.86232 \\
$\eta_y$ [-]           & & 26      & & 26      & 26      & 26      & & 26      & 26      \\
$\bar{y}_1$ [nm]       & & -0.1002 & & -1.0664 & -1.0664 & -1.5462 & & -1.0657 & -1.0651 \\
$\bar{y}_2$ [nm]       & & -0.8620 & & -0.8629 & -0.8629 & -1.3423 & & -0.8618 & -0.8624 \\
$\bar{z}_1$ [nm]       & & 0.4477  & & 0.3830  & 0.7830  & 0.4462  & & 0.5680  & 0.4467  \\
$\bar{z}_2$ [nm]       & & 0.4477  & & 0.3830  & 0.3830  & 0.4462  & & 0.5680  & 0.4467  \\
$\sigma_z$ [nm]        & & 0.06    & & 0.06    & 0.06    & 0.06    & & 0.06    & 0.06    \\
$\sigma_\text{st}$ [-] & & 150     & & 150     & 150     & 150     & & 150     & 150     \\
$\tilde{s}_{\text{st},0}$ [-] & & 0.10 & & 0.10 & 0.10 & 0.10 & & 0.10 & 0.10 \\
$k_\text{st}$ [kJ/mol]        & & 15   & & 15   & 15   & 15   & & 15   & 15   \\
$s_{\text{st},0}$ [-]         & & 24   & & 24   & 24   & 48   & & 24   & 48   \\
\bottomrule
\end{tabular*}
\label{tab:sstNa4}
\end{table}

\begin{table}[!htbp]
\caption{Values of the surface structure CV parameters for AcO$^-$ of layer 4.}
\centering
\begin{tabular*}{\textwidth}{@{\extracolsep{\fill}}lccccccccc}
\toprule
                   & & pure MeOH  & & \multicolumn{3}{c}{MeOH-PrOH} & & \multicolumn{2}{c}{MeOH-MeCN} \\
\si{\percent}      & & 100        & & 80-20      & 60-40   & 40-60  & &  75-25        &  50-50        \\
\midrule
$\nu_x$ [-]            & & 4       & & 4       & 4       & 6       & & 4       & 6       \\
$L_x$ [nm]             & & 2.48781 & & 2.48781 & 2.48781 & 3.73161 & & 2.48781 & 3.73161 \\
$\eta_x$ [-]           & & 12      & & 12      & 12      & 12      & & 12      & 12      \\
$\bar{x}_1$ [nm]       & & -0.9132 & & -0.9135 & -0.9135 & -1.5720 & & -0.9132 & -1.5700 \\
$\bar{x}_2$ [nm]       & & -0.7531 & & -0.7525 & -0.7525 & -1.4248 & & -0.7527 & -1.4215 \\
$\bar{x}_3$ [nm]       & & -0.6031 & & -0.6029 & -0.6029 & -1.2609 & & -0.6018 & -1.2589 \\
$\bar{x}_4$ [nm]       & & -0.4422 & & -0.4421 & -0.4421 & -1.1135 & & -0.4427 & -1.1106 \\
$\nu_y$ [-]            & & 3       & & 3       & 3       & 4       & & 3       & 4       \\
$L_y$ [nm]             & & 2.89733 & & 2.89733 & 2.89733 & 3.86232 & & 2.89733 & 3.86232 \\
$\eta_y$ [-]           & & 30      & & 30      & 30      & 30      & & 30      & 30      \\
$\bar{y}_1$ [nm]       & & -0.8343 & & -0.8348 & -0.8348 & -1.3146 & & -0.8332 & -0.8347 \\
$\bar{y}_2$ [nm]       & & -0.8343 & & -0.8348 & -0.8348 & -1.5738 & & -0.8332 & -1.0952 \\
$\bar{y}_3$ [nm]       & & -0.1281 & & -1.0946 & -1.0946 & -1.5738 & & -1.0934 & -1.0952 \\
$\bar{y}_4$ [nm]       & & -0.1281 & & -1.0946 & -1.0946 & -1.3146 & & -1.0934 & -0.8347 \\
$\bar{z}_1$ [nm]       & & 0.3817  & & 0.3167  & 0.3167  & 0.3826  & & 0.5020  & 0.3832  \\
$\bar{z}_2$ [nm]       & & 0.5153  & & 0.4502  & 0.4502  & 0.5159  & & 0.6360  & 0.5154  \\
$\bar{z}_3$ [nm]       & & 0.3817  & & 0.3167  & 0.3167  & 0.3826  & & 0.5020  & 0.3832  \\
$\bar{z}_4$ [nm]       & & 0.5153  & & 0.4502  & 0.4502  & 0.5159  & & 0.6360  & 0.5154  \\
$\sigma_z$ [nm]        & & 0.06    & & 0.06    & 0.06    & 0.06    & & 0.06    & 0.06    \\
$\sigma_\text{st}$ [-] & & 150     & & 150     & 150     & 150     & & 150     & 150     \\
$\tilde{s}_{\text{st},0}$ [-] & & 0.10 & & 0.10 & 0.10 & 0.10 & & 0.10 & 0.10 \\
$k_\text{st}$ [kJ/mol]        & & 15   & & 15   & 15   & 15   & & 15   & 15   \\
$s_{\text{st},0}$ [-]         & & 48   & & 48   & 48   & 96   & & 48   & 96   \\
\bottomrule
\end{tabular*}
\label{tab:sstAcO4}
\end{table}

\begin{table}[!htbp]
\caption{Values of the surface structure CV parameters for Na$^+$ of layer 1.}
\centering
\begin{tabular*}{\textwidth}{@{\extracolsep{\fill}}lccccccccc}
\toprule
                   & & pure MeOH  & & \multicolumn{3}{c}{MeOH-PrOH} & & \multicolumn{2}{c}{MeOH-MeCN} \\
\si{\percent}      & & 100        & & 80-20      & 60-40   & 40-60  & &  75-25        &  50-50        \\
\midrule
$\nu_x$ [-]            & & 4       & & 4       & 4       & 6       & & 4       & 6       \\
$L_x$ [nm]             & & 2.48781 & & 2.48781 & 2.48781 & 3.73161 & & 2.48781 & 3.73161 \\
$\eta_x$ [-]           & & 10      & & 10      & 10      & 10      & & 10      & 10      \\
$\bar{x}_1$ [nm]       & & -0.9195 & & -0.9193 & -0.9193 & -1.2510 & & -0.9180 & -1.2546 \\
$\bar{x}_2$ [nm]       & & -0.6083 & & -0.6092 & -0.6092 & -0.9389 & & -0.6084 & -0.9439 \\
$\nu_y$ [-]            & & 3       & & 3       & 3       & 4       & & 3       & 4       \\
$L_y$ [nm]             & & 2.89733 & & 2.89733 & 2.89733 & 3.86232 & & 2.89733 & 3.86232 \\
$\eta_y$ [-]           & & 26      & & 26      & 26      & 26      & & 26      & 26      \\
$\bar{y}_1$ [nm]       & & -1.0600 & & -1.0598 & -1.0598 & -1.3512 & & -1.0596 & -0.8683 \\
$\bar{y}_2$ [nm]       & & -0.8655 & & -0.8682 & -0.8682 & -0.5760 & & -0.8676 & -0.0939 \\
$\bar{z}_1$ [nm]       & & -0.4536 & & -0.5187 & -0.5187 & -0.4531 & & -0.3329 & -0.4538 \\
$\bar{z}_2$ [nm]       & & -0.4536 & & -0.5187 & -0.5187 & -0.4531 & & -0.3329 & -0.4538 \\
$\sigma_z$ [nm]        & & 0.06    & & 0.06    & 0.06    & 0.06    & & 0.06    & 0.06    \\
$\sigma_\text{st}$ [-] & & 150     & & 150     & 150     & 150     & & 150     & 150     \\
$\tilde{s}_{\text{st},0}$ [-] & & 0.10 & & 0.10 & 0.10 & 0.10 & & 0.10 & 0.10 \\
$k_\text{st}$ [kJ/mol]        & & 15   & & 15   & 15   & 15   & & 15   & 15   \\
$s_{\text{st},0}$ [-]         & & 24   & & 24   & 24   & 48   & & 24   & 48   \\
\bottomrule
\end{tabular*}
\label{tab:sstNa1}
\end{table}

\begin{table}[!htbp]
\caption{Values of the surface structure CV parameters for AcO$^-$ of layer 1.}
\centering
\begin{tabular*}{\textwidth}{@{\extracolsep{\fill}}lccccccccc}
\toprule
                   & & pure MeOH  & & \multicolumn{3}{c}{MeOH-PrOH} & & \multicolumn{2}{c}{MeOH-MeCN} \\
\si{\percent}      & & 100        & & 80-20      & 60-40   & 40-60  & &  75-25        &  50-50        \\
\midrule
$\nu_x$ [-]            & & 4       & & 4       & 4       & 6       & & 4       & 6       \\
$L_x$ [nm]             & & 2.48781 & & 2.48781 & 2.48781 & 3.73161 & & 2.48781 & 3.73161 \\
$\eta_x$ [-]           & & 12      & & 12      & 12      & 12      & & 12      & 12      \\
$\bar{x}_1$ [nm]       & & -1.0043 & & -1.0050 & -1.0050 & -1.1659 & & -1.0024 & -1.1651 \\
$\bar{x}_2$ [nm]       & & -0.8331 & & -0.8347 & -0.8347 & -1.0258 & & -0.8334 & -1.0274 \\
$\bar{x}_3$ [nm]       & & -0.6922 & & -0.6924 & -0.6924 & -0.8546 & & -0.6917 & -0.8600 \\
$\bar{x}_4$ [nm]       & & -0.5235 & & -0.5235 & -0.5235 & -0.7137 & & -0.5224 & -0.7181 \\
$\nu_y$ [-]            & & 3       & & 3       & 3       & 4       & & 3       & 4       \\
$L_y$ [nm]             & & 2.89733 & & 2.89733 & 2.89733 & 3.86232 & & 2.89733 & 3.86232 \\
$\eta_y$ [-]           & & 30      & & 30      & 30      & 30      & & 30      & 30      \\
$\bar{y}_1$ [nm]       & & -0.8266 & & -0.8249 & -0.8249 & -0.6206 & & -0.8264 & -0.1365 \\
$\bar{y}_2$ [nm]       & & -0.8266 & & -0.8249 & -0.8249 & -1.3073 & & -0.8264 & -0.8257 \\
$\bar{y}_3$ [nm]       & & -0.1371 & & -0.1350 & -0.1350 & -1.3073 & & -1.1017 & -0.8257 \\
$\bar{y}_4$ [nm]       & & -0.1371 & & -0.1350 & -0.1350 & -0.6206 & & -1.1017 & -0.1365 \\
$\bar{z}_1$ [nm]       & & -0.3953 & & -0.4573 & -0.4573 & -0.3973 & & -0.2750 & -0.3971 \\
$\bar{z}_2$ [nm]       & & -0.5096 & & -0.5739 & -0.5739 & -0.5090 & & -0.3897 & -0.5096 \\
$\bar{z}_3$ [nm]       & & -0.3953 & & -0.4573 & -0.4573 & -0.3973 & & -0.2750 & -0.3971 \\
$\bar{z}_4$ [nm]       & & -0.5096 & & -0.5739 & -0.5739 & -0.5090 & & -0.3897 & -0.5096 \\
$\sigma_z$ [nm]        & & 0.06    & & 0.06    & 0.06    & 0.06    & & 0.06    & 0.06    \\
$\sigma_\text{st}$ [-] & & 150     & & 150     & 150     & 150     & & 150     & 150     \\
$\tilde{s}_{\text{st},0}$ [-] & & 0.10 & & 0.10 & 0.10 & 0.10 & & 0.10 & 0.10 \\
$k_\text{st}$ [kJ/mol]        & & 15   & & 15   & 15   & 15   & & 15   & 15   \\
$s_{\text{st},0}$ [-]         & & 48   & & 48   & 48   & 96   & & 48   & 96   \\
\bottomrule
\end{tabular*}
\label{tab:sstAcO1}
\end{table}

\subsection{Biased CVs} \label{sec:S4.3} 

Since the growth and dissolution of Na$^+$ as well as AcO$^-$ are rare events, we need to to enhance the process by applying WTMetaD to overcome the timescale limitations and obtain sufficient growth and dissolution events. 
Only with sufficient sampling can we reliably compute the energy difference between grown and dissolved dimeric unit states. 
The slow degrees of freedom for the ions of organic salts are the same ones as for organic molecules, which we have investigated in our previous work \cite{Bjelobrk2021}. These are the diffusion of the solute to the kink site,
the desorption of the solvent from the particular site, and the partial desolvation and adsorption of the solute at the kink site. We therefore use CVs with the same functional form as the ones reported in Reference \citenum{Bjelobrk2021}.

The WTMetaD bias potential is introduced for each ion through the particular biased CV, $s_{\text{b},\text{ion}}$. Each $s_{\text{b},\text{ion}}$ is a set of functions, which are linear combinations of the local density of the particular ion and the local density of solvent and antisolvent molecules at the corresponding kink site of the ion\cite{Bjelobrk2021}.
The densities are defined as sums of Gaussian like bell curve terms; for each ion
\begin{equation}
s_{\text{s},\text{ion}} = \sum_i \exp \left( -\frac{|\mathbf{r}_i - \mathbf{r}_{\text{s},\text{ion}}|^2}{2\sigma_{\text{s},\text{ion}}^2} \right),
\end{equation}
and for the corresponding solvent and antisolvent
\begin{equation}
s_{\text{l},\text{ion}} = \sum_j \exp \left( -\frac{|\mathbf{r}_j - \mathbf{r}_{\text{l},\text{ion}}|^2}{2\sigma_{\text{l},\text{ion}}^2} \right).
\end{equation}
$\mathbf{r}_i$ and $\mathbf{r}_j$ are the positions of the ion $i$ and solvent or antisolvent $j$. 
$\mathbf{r}_{\text{s},\text{ion}}$ is the location of the ion's adsorption site, and identical with its crystal lattice position at the kink site. $\mathbf{r}_{\text{l},\text{ion}}$ is the adsorption site of the solvent or antisolvent at the solvated kink site.
$\mathbf{r}_{\text{s},\text{ion}}$ and $\mathbf{r}_{\text{l},\text{ion}}$ are for the investigated NaOAc kink sites not identical positions. 
$\sigma_{\text{s},\text{ion}}$ and $\sigma_{\text{l},\text{ion}}$ define the width of each Gaussian like bell curve.

$s_{\text{s},\text{ion}}$ captures the slow coordinate of diffusion and adsorption/desorption of the ion. $s_{\text{l},\text{ion}}$ captures the slow coordinate of desolvation/solvation of the kink site. These functions are combined into the biased CV
\begin{equation}
s_{\text{b},\text{ion}} = w_{\text{s},\text{ion}} (s_{\text{s},\text{ion}}^{\chi_{\text{s1},\text{ion}}} + s_{\text{s},\text{ion}}^{\chi_{\text{s2},\text{ion}}}) + w_{\text{l},\text{ion}} s_{\text{l},\text{ion}}^{\chi_{\text{l},\text{ion}}}.
\end{equation}
$w_{\text{s},\text{ion}}$ and $w_{\text{l},\text{ion}}$ are linear weights of each local density. $s_{\text{b},\text{ion}}$ has the same functional form as the one reported in our previous work \cite{Bjelobrk2021,Mendels2018a}.
The performance of the WTMetaD sampling is improved significantly by mapping\cite{Bjelobrk2019,Rizzi2020} the local densities as follows
\begin{equation}
s_{\text{s},\text{ion}} \rightarrow s_{\text{s},\text{ion}}^{\chi_{\text{s1},\text{ion}}} + s_{\text{s},\text{ion}}^{\chi_{\text{s2},\text{ion}}},
\end{equation}
and
\begin{equation}
s_{\text{l},\text{ion}} \rightarrow s_{\text{l},\text{ion}}^{\chi_{\text{l},\text{ion}}},
\end{equation}
with the scalar positive exponents, $\chi_\text{s1} < 1$, $\chi_\text{s2} > 1$ and $\chi_\text{l} < 1$. The mapping of the density functions allows us firstly, to obtain for the FES, $F(s_{\text{b},\text{ion}})$, broad local energy minima basins of equal widths. This enables us to use a larger Gaussian width of the bias potentials, $\sigma_W$, in the WTMetaD simulations, which consequently accelerates convergence. Secondly, the mapping allows us to push $s_{\text{s},\text{ion}}$ and $s_{\text{l},\text{ion}}$ considerably away from values of zero, at which the WTMetaD sampling performance can be otherwise disturbed. More details can be found in Reference \citenum{Bjelobrk2019}.
Without mapping the density functions appropriately, the biased simulations will not converge.

For the organic compounds involved in the biased CVs of Na$^+$, $s_{\text{b},\text{Na}+}$, and AcO$^-$, $s_{\text{b},\text{AcO}-}$, following ion and molecule centers were defined;
\begin{itemize}
\item[$\diamond$] AcO$^-$: center of mass of oxygen atoms (O1 and O2),
\item[$\diamond$] MeOH: oxygen atom (OM1),
\item[$\diamond$] PrOH: oxygen atom (OP1),
\item[$\diamond$] MeCN: nitrogen atom (NA1).
\end{itemize}
From unbiased simulations, one can obtain the orientations and positions of the ions as well as solvent and antisolvent molecules at the kink site. The alcohol oxygens and acetonitrile nitrogen share the same adsorption sites. However, the positions of these solvent and antisolvent adsorption sites differ for Na$^+$ and AcO$^-$.

Contour lines of $s_{\text{s},\text{Na+}}$ and $s_{\text{s},\text{AcO-}}$ are shown in Figures \ref{fig:adsorptionsites}a and \ref{fig:adsorptionsites}b respectively.

In simulations involving WTMetaD, it is beneficial to introduce lower and upper wall potentials for the biased CVs to improve the sampling performance. We therefore have used following wall potentials for the biased CVs of each ion
\begin{equation}
V_{\text{s},\text{ion}} =
\begin{cases}
k_\text{s,l} ((s_\text{s}^{\chi_\text{s1}}+s_\text{s}^{\chi_\text{s2}}) - s_\text{s,l})^2, & \text{if}\ (s_\text{s}^{\chi_\text{s1}}+s_\text{s}^{\chi_\text{s2}}) < s_\text{s,l}, \\
k_\text{s,u} ((s_\text{s}^{\chi_\text{s1}}+s_\text{s}^{\chi_\text{s2}}) - s_\text{s,u})^2, & \text{if}\ (s_\text{s}^{\chi_\text{s1}}+s_\text{s}^{\chi_\text{s2}}) > s_\text{s,u}, \\
0, & \text{else},
\end{cases}
\end{equation}
and
\begin{equation}
V_{\text{l},\text{ion}} =
\begin{cases}
k_\text{l,l} (s_\text{l}^{\chi_\text{l}} - s_\text{l,l})^2, & \text{if}\ s_\text{l}^{\chi_\text{l}} < s_\text{l,l}, \\
k_\text{l,u} (s_\text{l}^{\chi_\text{l}} - s_\text{l,u})^2, & \text{if}\ s_\text{l}^{\chi_\text{l}} > s_\text{l,u}, \\
0, & \text{else}.
\end{cases}
\end{equation}
$k_\text{s,l}$, $k_\text{s,u}$, $k_\text{l,l}$, and $k_\text{l,u}$ are the force constants and $s_\text{s,l}$, $s_\text{s,u}$, $s_\text{l,l}$, and $s_\text{l,u}$ are the thresholds below and above which the potentials are active. Lower walls are necessary to prevent the biased simulations from getting stuck at CV values of zero. Higher walls prevent the biased system from visiting non-physical states of excessive agglomeration of ions or solvent and antisolvent molecules at the kink site.

The parameter values used for the biased CVs are presented in Table \ref{tab:biasedCVs} and the WTMetaD parameters are listed in Table \ref{tab:WTMetaD}.

\begin{table}
\caption{Values of the biased CVs used for Na$^+$ and AcO$^-$ kink growth and dissolution sampling.}
\centering
\begin{tabular*}{\textwidth}{@{\extracolsep{\fill}}llccccccccc}
\toprule
&                   & & pure MeOH  & & \multicolumn{3}{c}{MeOH-PrOH} & & \multicolumn{2}{c}{MeOH-MeCN} \\
& \si{\percent}     & & 100        & & 80-20      & 60-40   & 40-60  & &  75-25        &  50-50        \\
\midrule
\multirow{7}{*}{$s_{\text{s},\text{Na}^+}$}
& $r_\text{s}^{(x)}$ [nm] & & 1.4584 & & 1.4576 & 1.4567 & 1.6569 & & 1.4587 & 1.3433 \\
& $r_\text{s}^{(y)}$ [nm] & & 1.3558 & & 1.3572 & 1.5447 & 1.3601 & & 1.3575 & 2.0351 \\
& $r_\text{s}^{(z)}$ [nm] & & 0.7506 & & 0.6860 & 0.7562 & 0.7514 & & 0.8713 & 0.7506 \\
& $\sigma_\text{s}$ [-]   & & 0.16   & & 0.16   & 0.16   & 0.16   & & 0.16   & 0.16   \\
& $\chi_\text{s}$ [-]     & & 0.3    & & 0.3    & 0.3    & 0.3    & & 0.3    & 0.3    \\
& $\chi_\text{s}$ [-]     & & 3      & & 3      & 3      & 3      & & 3      & 3      \\
& $w_\text{s}$ [-]        & & 0.5    & & 0.5    & 0.5    & 0.5    & & 0.5    & 0.5    \\
& $k_\text{s,l}$ [kJ/mol] & & 25     & & 25     & 25     & 25     & & 25     & 25     \\
& $k_\text{s,u}$ [kJ/mol] & & 25     & & 25     & 25     & 25     & & 25     & 25     \\
& $s_\text{s,l}$ [-]      & & 0.02   & & 0.02   & 0.02   & 0.02   & & 0.02   & 0.02   \\
& $s_\text{s,u}$ [-]      & & 2.08   & & 2.08   & 2.08   & 2.08   & & 2.08   & 2.08   \\
\midrule
\multirow{6}{*}{$s_{\text{l},\text{Na}^+}$}
& $r_\text{l}^{(x)}$ [nm] & & 1.3886 & & 1.3845 & 1.3845 & 1.7300 & & 1.3880 & 1.4240 \\
& $r_\text{l}^{(y)}$ [nm] & & 1.5250 & & 1.4950 & 1.4000 & 1.4750 & & 1.5020 & 1.8550 \\
& $r_\text{l}^{(z)}$ [nm] & & 0.6863 & & 0.6150 & 0.6860 & 0.6800 & & 0.8000 & 0.6840 \\
& $\sigma_\text{l}$ [-]   & & 0.05   & & 0.05   & 0.05   & 0.05   & & 0.05   & 0.05   \\
& $\chi_\text{l}$ [-]     & & 0.6    & & 0.6    & 0.6    & 0.6    & & 0.6    & 0.6    \\
& $w_\text{l}$ [-]        & & -0.25  & & -0.25  & -0.25  & -0.25  & & -0.25  & -0.25  \\
& $k_\text{l,l}$ [kJ/mol] & & 25     & & 25     & 25     & 25     & & 25     & 25     \\
& $k_\text{l,u}$ [kJ/mol] & & 25     & & 25     & 25     & 25     & & 25     & 25     \\
& $s_\text{l,l}$ [-]      & & 0.02   & & 0.02   & 0.02   & 0.02   & & 0.02   & 0.02   \\
& $s_\text{l,u}$ [-]      & & 1.05   & & 1.05   & 1.05   & 1.05   & & 1.05   & 1.05   \\
\midrule\midrule
\multirow{7}{*}{$s_{\text{s},\text{AcO}^-}$}
& $r_\text{s}^{(x)}$ [nm] & & 1.4571 & & 1.4576 & 1.4559 & 1.6577 & & 1.4583 & 1.3449 \\
& $r_\text{s}^{(y)}$ [nm] & & 1.5908 & & 1.5911 & 1.3103 & 1.5981 & & 1.5916 & 1.7992 \\
& $r_\text{s}^{(z)}$ [nm] & & 0.7442 & & 0.6794 & 0.7496 & 0.7484 & & 0.8639 & 0.7476 \\
& $\sigma_\text{s}$ [-]   & & 0.16   & & 0.16   & 0.16   & 0.16   & & 0.16   & 0.16   \\
& $\chi_\text{s}$ [-]     & & 0.3    & & 0.3    & 0.3    & 0.3    & & 0.3    & 0.3    \\
& $\chi_\text{s}$ [-]     & & 3      & & 3      & 3      & 3      & & 3      & 3      \\
& $w_\text{s}$ [-]        & & 0.5    & & 0.5    & 0.5    & 0.5    & & 0.5    & 0.5    \\ 
& $k_\text{s,l}$ [kJ/mol] & & 25     & & 25     & 25     & 25     & & 25     & 25     \\
& $k_\text{s,u}$ [kJ/mol] & & 25     & & 25     & 25     & 25     & & 25     & 25     \\
& $s_\text{s,l}$ [-]      & & 0.02   & & 0.02   & 0.02   & 0.02   & & 0.02   & 0.02   \\
& $s_\text{s,u}$ [-]      & & 2.08   & & 2.08   & 2.08   & 2.08   & & 2.08   & 2.08   \\
\midrule
\multirow{6}{*}{$s_{\text{l},\text{AcO}^-}$}
& $r_\text{l}^{(x)}$ [nm] & & 1.3780 & & 1.3800 & 1.3770 & 1.7337 & & 1.3770 & 1.4240 \\
& $r_\text{l}^{(y)}$ [nm] & & 1.6380 & & 1.6450 & 1.2650 & 1.6467 & & 1.6420 & 1.7410 \\
& $r_\text{l}^{(z)}$ [nm] & & 0.6838 & & 0.6100 & 0.6850 & 0.6800 & & 0.7970 & 0.6800 \\
& $\sigma_\text{l}$ [-]   & & 0.10   & & 0.10   & 0.10   & 0.10   & & 0.10   & 0.10   \\
& $\chi_\text{l}$ [-]     & & 0.5    & & 0.5    & 0.5    & 0.5    & & 0.5    & 0.5    \\
& $w_\text{l}$ [-]        & & -1     & & -1     & -1     & -1     & & -1     & -1     \\
& $s_\text{l,l}$ [-]      & & 0.02   & & 0.02   & 0.02   & 0.02   & & 0.02   & 0.02   \\
& $s_\text{l,u}$ [-]      & & 1.05   & & 1.05   & 1.05   & 1.05   & & 1.05   & 1.05   \\
\hline   
\end{tabular*}
\label{tab:biasedCVs}
\end{table}

\begin{table}[!htbp]
\caption{Well-tempered Metadynamics parameter values used for the Na$^+$ and AcO$^-$ growth and dissolution simulations.} \centering
\begin{tabular*}{\textwidth}{@{\extracolsep{\fill}}lcccc}
\toprule
                        & & Na$^+$    & & AcO$^-$    \\
\midrule
$W$ [kJ/mol]            & & 0.2       & & 0.2        \\
$\sigma_W$ [-]          & & 0.06      & & 0.06       \\
$\gamma$ [-]            & & 8         & & 10         \\
$\tau$ [ps]             & & 1         & & 1          \\
$\Delta s_\text{b}$ [-] & & 0.02      & & 0.02       \\
\bottomrule
\end{tabular*}
\label{tab:WTMetaD}
\end{table}

\subsection{Crystallinity CVs} \label{sec:S4.4} 

To compute the energy difference between crystalline and dissolved kink site states, one requires a measure of the crystallinity of the kink site.
The biased CVs, which are designed to capture the slow degrees of freedom of the kink growth and dissolution process, do not define the crystallinity of the kink site.
We therefore have to define another set of CVs, which do capture the crystallinity. We define a set of logistic functions, which are of same functional form as the spherical adsorption site CVs discussed in Section \ref{sec:S4.1}, for a particular atom of NaOAc
\begin{equation}
s_{\text{c},\text{atom}} = \sum_i \left(1 - \frac{1}{1+\exp(-\sigma_\text{c}(|\mathbf{r}_{\text{c},i}-\mathbf{\bar{r}}_\text{c}|-d_\text{c}))}\right),
\end{equation}
with $\sigma_\text{c}$ defining the width and $d_\text{c}$ the position of the switching function. Vector $\mathbf{r}_{\text{c},i}$ is the atom position of the $i$-th ion and $\mathbf{\bar{r}}_{\text{c}}$ is the atom's position in the crystalline lattice at the kink site.

Since Na$^+$ is spherical, we need one switching function to define its crystallinity, i.e. CV $s_{\text{c},\text{Na}+}$. For AcO$^-$ it is necessary to use two switching functions to account for the ion's orientation as well. For this purpose we take AcO$^-$'s atom C1 for crystallinity CV $s_{\text{c1},\text{AcO}-}$ and the center of mass of AcO$^-$'s oxygen atoms O1 and O2 for crystallinity CV $s_{\text{c1},\text{AcO}-}$.

$\mathbf{\bar{r}}_{\text{c}}$ are easily extracted from unbiased simulations where the kink site is grown. 
$d_\text{c}$ and $\sigma_\text{c}$ can be interpreted as the position and width of the region of transition, in which the ion transforms from fully dissolved to crystalline.
We need to define at which distance from the kink site the ion is considered dissolved and where the ion is considered crystalline. In between lies the region where the ion is in neither state but in the region of transition.

We check the distance from the kink site where the ion is fully solvated and assign values of the crystallinity CV switching functions as zero (i.e. where the ions are fully solvated).
We assign CV values of one to the positions, where the ion is fully crystalline in the same way as in our previous work\cite{Bjelobrk2019} (we include also the amplitude of the lattice vibrations to the region, where the ion is crystalline). The values between zero and one are considered to belong to values of the transition region.

All crystallinity CVs parameter values used for the reweighting process (discussed in Section \ref{sec:S5}) are shown in Table \ref{tab:crystallinityCVs}.

\begin{table}
\caption{Values of the crystallinity CVs used in the reweighting.}
\centering
\begin{tabular*}{\textwidth}{@{\extracolsep{\fill}}llccccccccc}
\toprule
&                   & & pure MeOH  & & \multicolumn{3}{c}{MeOH-PrOH} & & \multicolumn{2}{c}{MeOH-MeCN} \\
& \si{\percent}     & & 100        & & 80-20      & 60-40   & 40-60  & &  75-25        &  50-50        \\
\midrule
\multirow{5}{*}{$s_{\text{c},\text{Na}^+}$}
& $r_\text{c}^{(x)}$ [nm]  & & 1.4584  & & 1.4576  & 1.4567  & 1.6569  & & 1.4587  & 1.3433  \\
& $r_\text{c}^{(x)}$ [nm]  & & 1.3558  & & 1.3572  & 1.5447  & 1.3601  & & 1.3575  & 2.0351  \\
& $r_\text{c}^{(x)}$ [nm]  & & 0.7506  & & 0.6860  & 0.7562  & 0.7514  & & 0.8713  & 0.7506  \\
& $\sigma_\text{c}$ [-]    & & 65      & & 65      & 65      & 65      & & 65      & 65      \\
& $d_\text{c}$ [nm]        & & 0.15    & & 0.15    & 0.15    & 0.15    & & 0.15    & 0.15    \\
\midrule\midrule
\multirow{5}{*}{$s_{\text{c1},\text{AcO}^-}$}
& $r_\text{c1}^{(x)}$ [nm] & & 1.4277  & & 1.4269  & 1.4260  & 1.6877  & & 1.4280  & 1.3741  \\
& $r_\text{c1}^{(y)}$ [nm] & & 1.8113  & & 1.8127  & 1.0892  & 1.8156  & & 1.8130  & 1.5796  \\
& $r_\text{c1}^{(z)}$ [nm] & & 0.7789  & & 0.7143  & 0.7845  & 0.7797  & & 0.8996  & 0.7789  \\
& $\sigma_\text{c1}$ [-]   & & 80      & & 80      & 80      & 80      & & 80      & 80      \\
& $d_\text{c1}$ [nm]       & & 0.16    & & 0.16    & 0.16    & 0.16    & & 0.16    & 0.16    \\
\midrule
\multirow{5}{*}{$s_{\text{c2},\text{AcO}^-}$}
& $r_\text{c2}^{(x)}$ [nm] & & 1.4084  & & 1.4076  & 1.4067  & 1.7069  & & 1.4087  & 1.3933  \\
& $r_\text{c2}^{(y)}$ [nm] & & 1.6738  & & 1.6752  & 1.2267  & 1.6781  & & 1.6755  & 1.7171  \\
& $r_\text{c2}^{(z)}$ [nm] & & 0.7506  & & 0.6860  & 0.7562  & 0.7514  & & 0.8713  & 0.7506  \\
& $\sigma_\text{c2}$ [-]   & & 65      & & 65      & 65      & 65      & & 65      & 65      \\
& $d_\text{c2}$ [nm]       & & 0.15    & & 0.15    & 0.15    & 0.15    & & 0.15    & 0.15    \\
\bottomrule
\end{tabular*}
\label{tab:crystallinityCVs}
\end{table}

\section{Energy differences and WTMetaD convergence performance} \label{sec:S5}

For any enhanced sampling simulation, an estimate of the FES convergence performance is of key interest. 
Here we shall discuss the Na$^+$ and AcO$^-$ growth and dissolution simulations at kink sites, which were biased with WTMetaD, for the representative case of pure MeOH solutions.

\begin{figure}[!htbp]
\begin{center}
\includegraphics[width=\textwidth]{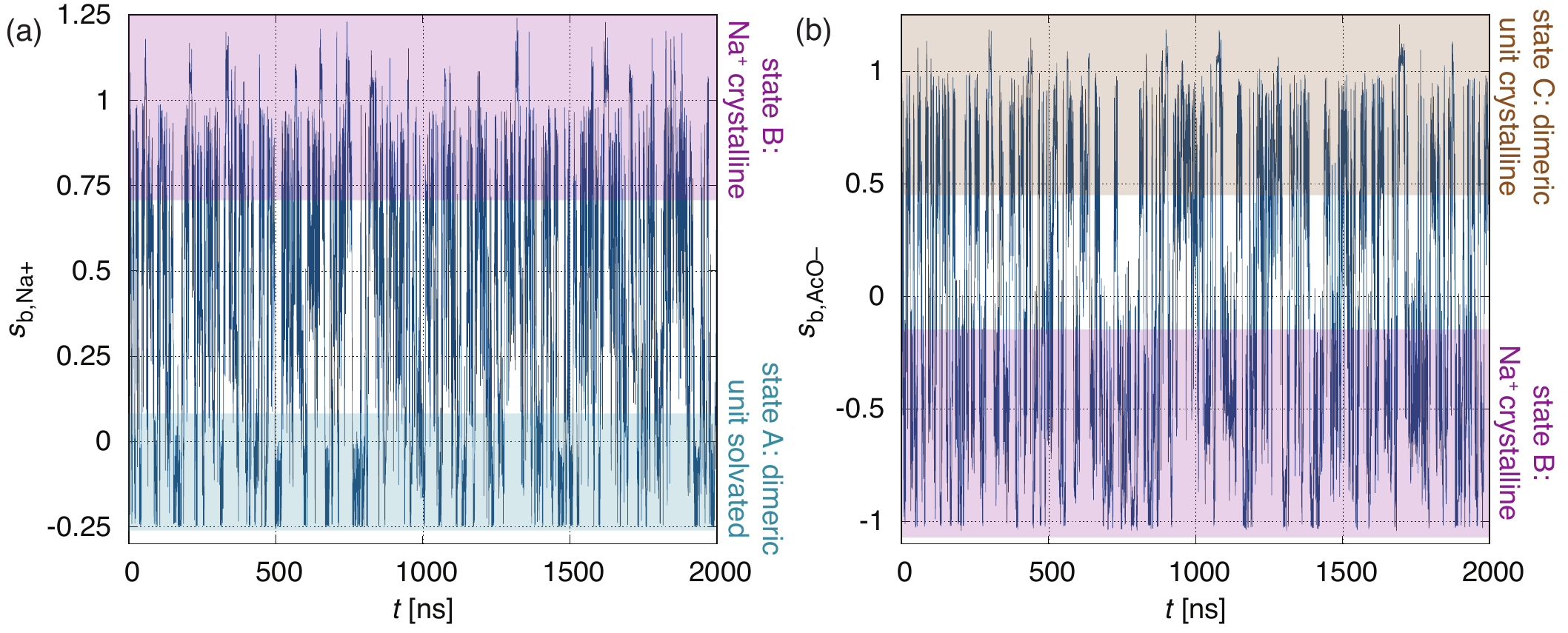}
\end{center}
\caption{Time trajectories of the biased CVs for each ion in pure MeOH solution for a mole fraction of $\chi = 0.0253$. a) Biased CV of Na$^+$, $s_{\text{b},\text{Na}+}$, vs. time; state A, where the biased dimeric unit is completely solvated, is shaded in blue and the state B, where Na$^+$ is crystalline in the dimeric unit, is shaded in purple. b) Biased CV of AcO$^-$, $s_{\text{b},\text{AcO}-}$, vs. time; state B is again colored in purple and state C, where the biased dimeric unit is crystalline, is colored in brown.} \label{fig:trajectories}
\end{figure}

For good WTMetaD sampling, the biased CV should transition between the different states without getting stuck or needing longer and longer periods between transitioning, which otherwise would indicate a hysteresis. Figures \ref{fig:trajectories}a and \ref{fig:trajectories}b show typical simulation runs for Na$^+$ and AcO$^-$ respectively at a solute mole fraction of $\chi = 0.0253$. Within 2 $\mu$s, the system transitions many times between grown and dissolved Na$^+$, and the same can be observed for AcO$^-$. No hysteresis was seen in any of the reported simulations.

Frequent transitioning between states is necessary but not sufficient for good convergence performance. The energy difference, $\Delta F$, between crystalline and dissolved states needs to converge over time and not fluctuate with an amplitude above the energy value required to resolve the energetic differences. In this case it is necessary to have an accuracy of $\Delta F$ in dependence of the solute mole fraction $\chi$ at a level that enables us to distinguish the solubilities for the different solvent/antisolvent mixtures.

\begin{figure}[!htbp]
\begin{center}
\includegraphics[width=\textwidth]{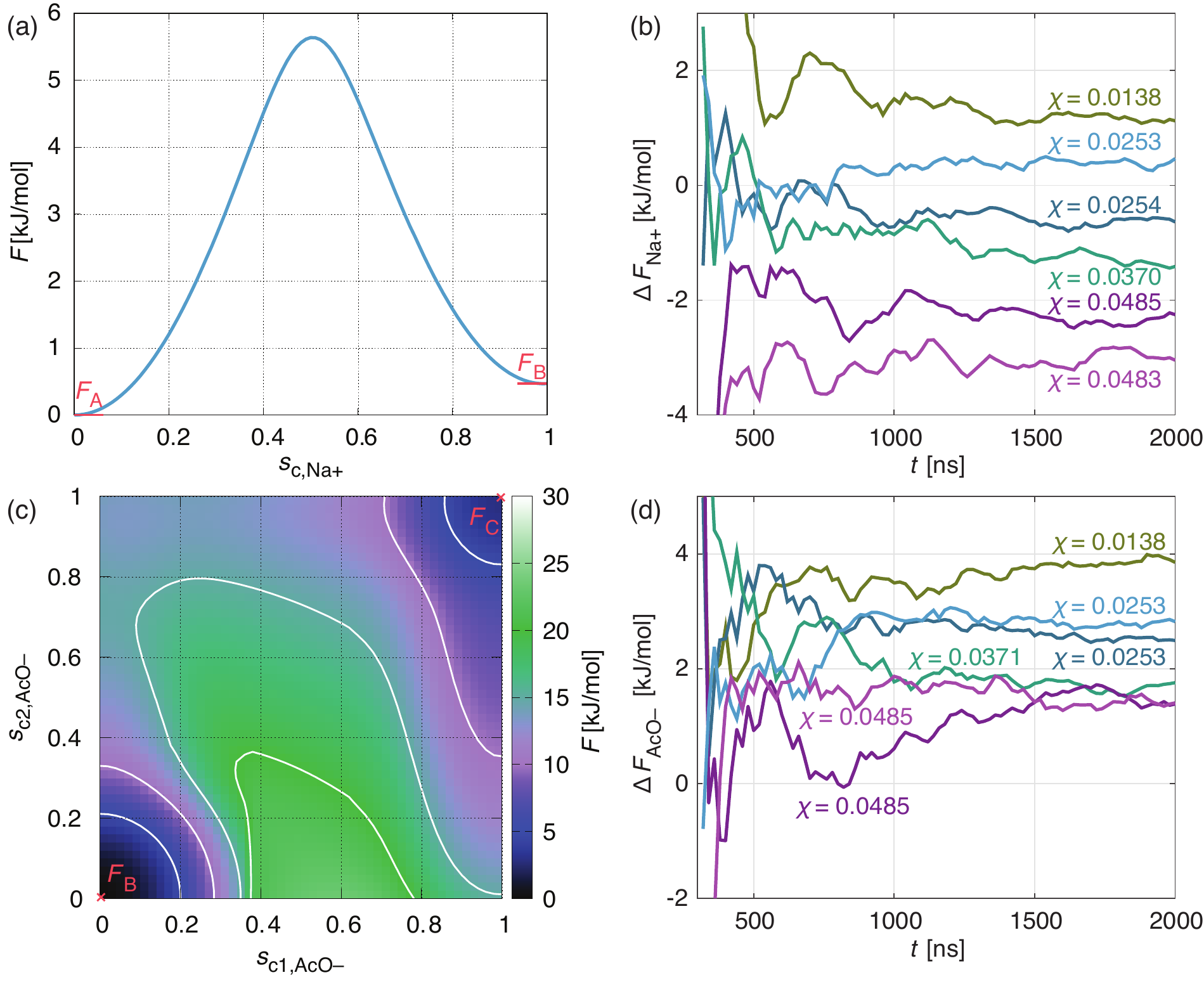}
\end{center}
\caption{a) FES, $F$, in dependence of the crystallinity CV, $s_{\text{c},\text{Na}+}$, obtained from the Na$^+$ growth and dissolution sampling in pure MeOH solution at a solute mole fraction of $\chi = 0.0253$. The difference between the energy values of the crystalline and dissolved kink site states, $F_\text{B}$ and $F_\text{A}$, yields $\Delta F_{\text{Na}+}$.
b) Time evolution of $\Delta F_{\text{Na}+}$ for all six reported pure MeOH solution simulation runs performed at the specified solute mole fractions.
c) $F$ in dependence of the crystallinity CVs, $s_{\text{c1},\text{AcO}-}$ and $s_{\text{c2},\text{AcO}-}$, obtained from the AcO$^-$ growth and dissolution sampling in pure MeOH solution at a solute mole fraction of $\chi = 0.0253$. The difference between the energy minima $F_\text{C}$ and $F_\text{B}$ gives $\Delta F_{\text{AcO}-}$.
d) Time evolution of $\Delta F_{\text{AcO}-}$ for all six reported pure MeOH solution simulations run at the specified solute mole fractions.
} \label{fig:convergenceanalysis}
\end{figure}

To obtain $\Delta F$, the FES in dependence of the kink site crystallinity needs to be computed. This is done by reweighting\cite{Tiwary2014} the biased simulations with the crystallinity CVs, which are discussed in Section \ref{sec:S4.4}. The FES in dependence of Na$^+$ crystallinity, $s_{\text{c},\text{Na}+}$, for a solute mole fraction of $\chi = 0.0253$ and simulation time $t = 2$ $\mu$s is presented in Figure \ref{fig:convergenceanalysis}a. As for all other biased simulations in the presented work, the first 300 ns of simulation were not used for the reweighting, since the WTMetaD bias potential changes significantly in this first time interval of the simulation. From the FES, the energy difference between crystallized and dissolved Na$^+$ can be obtained, $\Delta F_{\text{Na}+} = F_\text{B} - F_\text{A}$. The time evolution of $\Delta F_\text{Na+}$ is presented in Figure \ref{fig:convergenceanalysis}b for all reported Na$^+$ simulations in pure MeOH, run at the specified solute mole fractions. A reasonable convergence is attained around a simulation time of 1.2 $\mu$s.

Figure \ref{fig:convergenceanalysis}c shows the FES in dependence of the AcO$^-$ crystallinity CVs, $s_{\text{c1},\text{AcO}-}$ and $s_{\text{c2},\text{AcO}-}$. Again, the energy difference between crystalline and dissolved AcO$^-$ can be computed from the respective energy minima, i.e. $\Delta F_{\text{AcO}-} = F_\text{C} - F_\text{B}$. The time evolution of $\Delta F_{\text{AcO}-}$ for the specified solute mole fractions is shown in Figure \ref{fig:convergenceanalysis}. Again, reasonable convergence can be reached after simulation times of 1.2 $\mu$s.

As discussed in the main text, the energy difference for the dimeric unit can be obtained from summing the $\Delta F_\text{ion}$ of each simulation run pair of Na$^+$ and AcO$^-$, which was simulated at the same solute mole fraction, $\Delta F = \Delta F_{\text{Na}+} + \Delta F_{\text{AcO}-}$. The solubility can be identified at $\Delta F = 0$.

To average out energy difference fluctuations over time, the $\Delta F_\text{ion}$ reported in the main text and in Table \ref{tab:results} are the averaged values of $\Delta F_\text{ion}$ over the last 400 $\mu$s of simulation.

From the discussions above and results presented in the main text, we can conclude that the simulation setup has a convergence performance, that allows a reliable computation the solubilities of NaOAc in the reported solvent/antisolvent mixtures.

\section{Simulation results} \label{sec:S6}

The numeric values of the results shown in the results section of the main text are listed in Table \ref{tab:results}.
Several simulations were repeated to test the reproducibility of the method. The averages of the $\Delta F_\text{ion}$ values obtained from the repetitions were used for the linear regression.

\begin{table}[!htbp]
\caption{Values of mole fractions and energy differences for all performed biased kink growth and dissolution simulations.}
\centering
\begin{tabular*}{\textwidth}{@{\extracolsep{\fill}}lccccccccccc}
\toprule
 solution    & run       & & $\chi_{\text{Na}+}$ & $\Delta F_{\text{Na}+}$ & & $\chi_{\text{AcO}-}$ & $\Delta F_{\text{AcO}-}$ & & $\chi$ & $\Delta F$ \\
 composition & N$^\circ$ & & [-] & [kJ/mol] & & [-] & [kJ/mol] & & [-] & [kJ/mol]      \\
\midrule
\multirow{6}{*}{100\% MeOH}
& 1  & & 0.0138 &  1.1790 & & 0.0138 &  3.8549 & & 0.0138 &  5.0338  \\
& 2  & & 0.0254 & -0.6460 & & 0.0252 &  2.5242 & & 0.0253 &  1.8782  \\
& 3  & & 0.0253 &  0.3859 & & 0.0253 &  2.7714 & & 0.0253 &  3.1573  \\
& 4  & & 0.0370 & -1.2890 & & 0.0372 &  2.3237 & & 0.0371 &  0.3769  \\
& 5  & & 0.0485 & -2.3799 & & 0.0485 &  1.5283 & & 0.0485 & -0.8516  \\
& 6  & & 0.0483 & -2.9770 & & 0.0486 &  1.3946 & & 0.0485 & -1.5824  \\
\midrule
\multirow{7}{*}{80-20\% MeOH-PrOH}
& 1  & & 0.0177 &  0.8136 & & 0.0175 &  2.8134 & & 0.0176 &  3.6270  \\
& 2  & & 0.0174 & -0.0464 & & 0.0176 &  3.9107 & & 0.0175 &  3.8643  \\
& 3  & & 0.0245 & -0.6552 & & 0.0245 &  1.9105 & & 0.0245 &  1.2553  \\
& 4  & & 0.0311 & -0.1496 & & 0.0311 &  1.5104 & & 0.0311 &  1.3607  \\
& 5  & & 0.0311 & -1.3669 & & 0.0315 &  1.4911 & & 0.0313 &  0.1242  \\
& 6  & & 0.0377 & -1.4520 & & 0.0375 & -0.1456 & & 0.0376 & -1.5976  \\
& 7  & & 0.0374 & -1.9872 & & 0.0374 &  0.8069 & & 0.0374 & -1.1803  \\
\midrule
\multirow{7}{*}{60-40\% MeOH-PrOH}
& 1  & & 0.0112 &  1.9627 & & 0.0112 &  1.0170 & & 0.0112 &  2.9797  \\
& 2  & & 0.0117 &  3.3412 & & 0.0114 &  0.5339 & & 0.0116 &  3.8751  \\
& 3  & & 0.0191 &  1.2931 & & 0.0191 &  0.2854 & & 0.0191 &  1.5785  \\
& 4  & & 0.0192 &  0.4735 & & 0.0192 &  1.0516 & & 0.0192 &  1.5251  \\
& 5  & & 0.0272 & -0.3380 & & 0.0264 & -0.0116 & & 0.0268 & -0.3496  \\
& 6  & & 0.0344 & -2.6537 & & 0.0346 & -0.3976 & & 0.0345 & -3.0513  \\
& 7  & & 0.0340 & -2.7858 & & 0.0337 & -0.1634 & & 0.0339 & -2.9492  \\
\midrule
\multirow{7}{*}{40-60\% MeOH-PrOH}
& 1  & & 0.0065 &  3.9123 & & 0.0063 &  0.9312 & & 0.0064 &  4.8435  \\
& 2  & & 0.0105 &  3.1088 & & 0.0105 & -0.3701 & & 0.0105 &  2.7387  \\
& 3  & & 0.0141 &  2.1516 & & 0.0145 & -0.9524 & & 0.0143 &  1.1992  \\
& 4  & & 0.0140 &  1.9123 & & 0.0146 & -0.6265 & & 0.0143 &  1.2858  \\
& 5  & & 0.0182 &  1.6744 & & 0.0182 & -2.1048 & & 0.0182 & -0.4304  \\
& 6  & & 0.0183 &  1.2967 & & 0.0183 & -0.5738 & & 0.0183 &  0.7229  \\
& 7  & & 0.0215 &  0.2832 & & 0.0218 & -2.1561 & & 0.0217 & -1.8730  \\
\midrule
\multirow{7}{*}{75-25\% MeOH-MeCN}
& 1  & & 0.0092 &  2.0924 & & 0.0092 &  3.6488 & & 0.0092 &  5.7412  \\
& 2  & & 0.0158 &  0.8838 & & 0.0158 &  2.8862 & & 0.0158 &  3.7700  \\
& 3  & & 0.0227 & -1.2024 & & 0.0230 &  2.3339 & & 0.0229 &  1.1314  \\
& 4  & & 0.0299 & -1.6967 & & 0.0299 &  1.2920 & & 0.0299 & -0.4047  \\
& 5  & & 0.0299 & -1.6568 & & 0.0299 &  2.0634 & & 0.0299 &  0.4066  \\
& 6  & & 0.0365 & -2.5240 & & 0.0365 &  1.0251 & & 0.0365 & -1.4989  \\
& 7  & & 0.0365 & -3.4607 & & 0.0365 &  0.6874 & & 0.0365 & -2.7733  \\
\midrule
\multirow{6}{*}{50-50\% MeOH-MeCN}
& 1  & & 0.0022 &  2.9730 & & 0.0022 &  2.1770 & & 0.0022 &  5.1500 \\
& 2  & & 0.0051 &  1.9677 & & 0.0051 &  1.0203 & & 0.0051 &  2.9880 \\
& 3  & & 0.0051 &  2.3857 & & 0.0051 &  1.6836 & & 0.0051 &  4.0693 \\
& 4  & & 0.0081 &  0.9459 & & 0.0081 &  0.8041 & & 0.0081 &  1.7500 \\
& 5  & & 0.0110 &  0.2668 & & 0.0110 &  0.2713 & & 0.0110 &  0.5381 \\
& 6  & & 0.0137 &  0.5627 & & 0.0135 & -1.1132 & & 0.0136 & -0.5505 \\
\bottomrule
\end{tabular*}
\label{tab:results}
\end{table}

\section{Solubility dependency on sodium acetate force fields scaling factor} \label{sec:S7}

\begin{figure}[!htbp]
\begin{center}
\includegraphics[width=14cm]{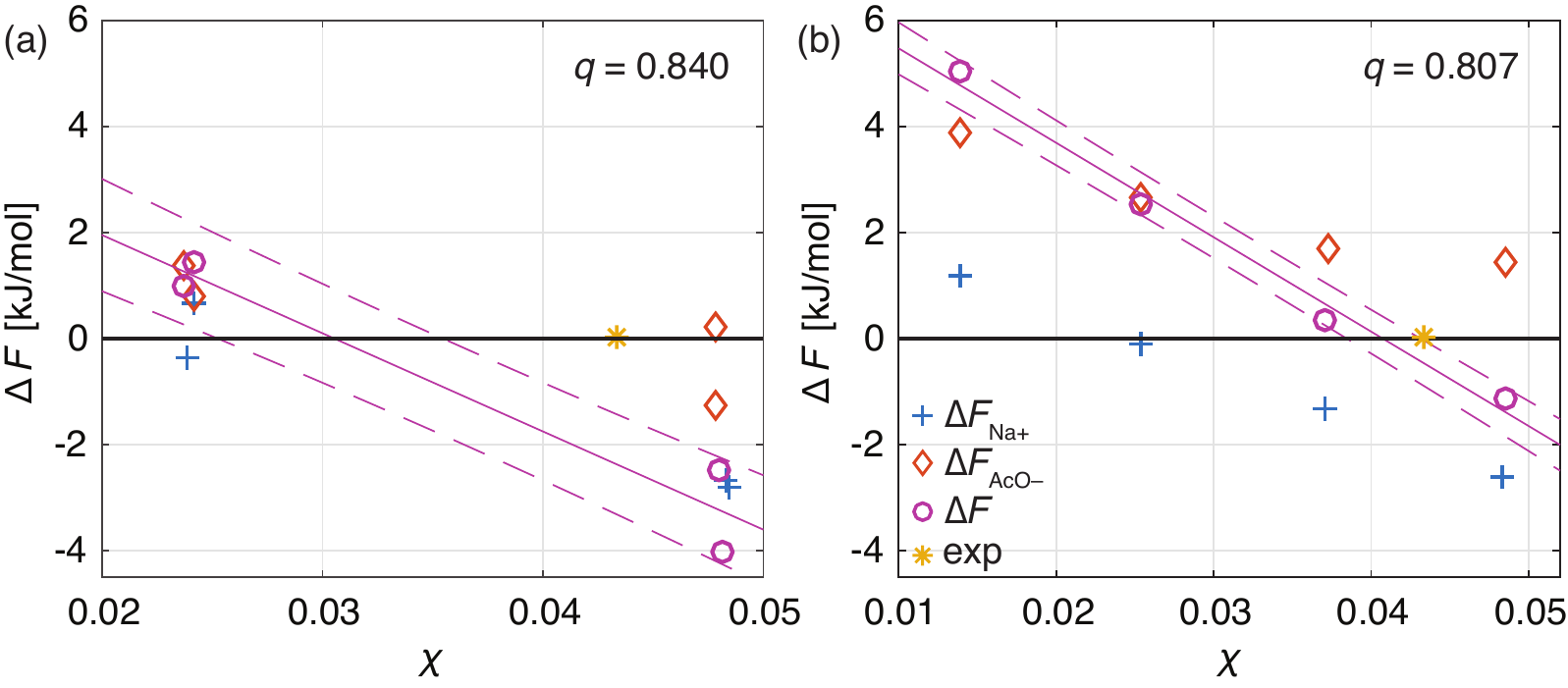}
\end{center}
\caption{Sampled energy differences in dependence of solute mole fraction $\chi$ for simulations using NaOAc force fields with different charge scaling factors: a) $q = 0.840$, and b) $q = 0.807$ (Figure 3a from main text). Energy differences are shown for Na$^+$, $\Delta F_{\text{Na}+}$ (blue crosses), AcO$^-$, $F_{\text{AcO}-}$ (orange diamonds), and their sums, $\Delta F = \Delta F_{\text{Na}+} + \Delta F_{\text{AcO}-}$ (purple circles). Linear fits of the $\Delta F$ data points are shown as solid lines and the lower and upper bounds of the standard deviations are shown as dashed lines. The yellow asterisks correspond to the experimental solubility of NaOAc in pure MeOH solution.} \label{fig:solubility_vs_q}
\end{figure}

As already mentioned in Section \ref{sec:S1.2}, the solubility is highly sensitive to the melting point of the crystalline compound, which is both true for real crystal systems as well as simulations. Consequently, starting off with a force field, which melting point considerably deviates from experiments, will lead to significant deviations between simulated and experimental solubility. We have run simulations with the NaOAc force fields \cite{Kashefolgheta2017} using a scaling factor of $q = 0.840$ and compared it to the runs of $q = 0.807$. The resulting $\Delta F$ and linear interpolation are presented in Figure \ref{fig:solubility_vs_q}, which we have obtained from biased simulations in pure MeOH solution. Figure \ref{fig:solubility_vs_q}a shows the simulation results for $q = 0.840$ and Figure \ref{fig:solubility_vs_q}b shows the results for $q = 0.807$.

Predicted solubilities were obtained from linear interpolation of the energy differences of the dimeric unit, $\Delta F$, in dependence of solute mole fraction, $\chi$, in the same fashion as in the main text.
Following values were obtained for the two different NaOAc force field systems:
\begin{itemize}
\item[$\diamond$] $\chi^*_\text{sim}(q = 0.840) = 0.031 \pm 0.005$,
\item[$\diamond$] $\chi^*_\text{sim}(q = 0.807) = 0.0407 \pm 0.0024$ (presented in main text).
\end{itemize}
Compared to the experimental solubility, $\chi_\text{exp}^* = 0.0440$, $\chi^*_\text{sim}(q = 0.840)$ has a considerable error of $-30 \%$; while $\chi^*_\text{sim}(q = 0.807)$ has an error of $-7.5 \%$, which is within the order of the simulation setup's standard deviation.
While $q = 0.807$ reproduces well the experimental melting point of NaOAc, $T_\text{m}^\text{exp} =$ 597 K, $q = 0.840$ leads to a melting point of $T_\text{m}^\text{sim} =$ 640 K, which is around 40 K above $T_\text{m}^\text{exp}$ (i.e. 7 \% relative error).
A small change in the columbic interaction parameters (i.e. small relative deviation in melting temperature) leads to a large change in solubility.
This comparison of results for different charge scaling factors clearly shows how important the correct melting temperature of force fields is for solubility predictions with MD simulations.

\end{document}